\documentclass[twocolumn,showpacs,amsmath,amssymb]{revtex4}

\usepackage{graphicx}
\usepackage{dcolumn}
\usepackage{bm}

\newcommand{\IGNORE}[1]{}
\newcommand{\R}         {\mathcal{R}}
\newcommand{\Btw}       {}
\newcommand{\units}     {\frac{\textrm{nm}}{\textrm{meV}}}
\newcommand{\invunits}  {\frac{\textrm{meV}}{\textrm{nm}}}

\begin{document}

\title{Interminiband Rabi oscillations in biased semiconductor superlattices}
\author{Pavel Abumov and D. W. L. Sprung}
\affiliation{
  Department of Physics and Astronomy, McMaster University\\
  Hamilton, Ontario L8S 4M1 Canada
}
\date{\today}

\begin{abstract}
Carrier dynamics at energy level anticrossings in biased
semiconductor superlattices, was studied in the time domain by
solving the time-dependent Schr\"odinger equation. The resonant
nature of interminiband Rabi oscillations has been explicitly
demonstrated to arise from interference of intrawell and Bloch
oscillations. We also report a simulation of direct Rabi
oscillations across three minibands, in the high field regime, due
to interaction between three strongly coupled minibands.
\end{abstract}



\maketitle


\section{\label{sec:intro}Introduction}

The rapid progress of solid-state electronics in recent years
has drawn attention to semiconductor (SC) superlattices (SL) as a
useful source of coherent electrons. The proposed applications
include microwave radiation (the so-called Bloch oscillator
systems~\cite{Bloch_Oscillator, Bloch_oscillator3,
Bloch_Oscillator2}), matter-wave interferometry~\cite{breid06_2} and
operational qubits for quantum computing~\cite{Zrenner02}. A better
understanding of fundamental types of carrier behaviour in biased
superlattices is essential for making further progress in this
field.

Under bias, semiconductor superlattices demonstrate some remarkable
quantum transport effects, such as resonant Zener tunneling (RZT)
and interminiband Rabi oscillations~\cite{Review_recent, Hone96,
Wacker_review02, Zhang04}. The phenomenon of interminiband Rabi
oscillations in SC SL is also known in the literature as excitonic
Rabi oscillations, Rabi flopping, periodic population swapping,
field-induced delocalization, oscillatory dipole, interwell
oscillations and Bloch-Zener oscillations. Previous
investigations~\cite{Rabi81, Voisin97, Rabi00} have covered many
aspects of Rabi oscillations in SC systems since their first
experimental observation in the early 1990's, by the laser
pump-and-probe technique~\cite{Schneider90}. In experiment, Rabi
oscillations occur as oscillating charge density dipoles that have
been observed both at low~\cite{Roskos92} and
room~\cite{Schneider90} temperatures. A typical system demonstrating
Rabi oscillations are SC quantum dots under pulsed resonant
excitation.

While a two-miniband model works well for shallow superlattices
(e.g. optical potentials), stronger potentials generally require a
more elaborate approach. In the past few years some authors have
employed more powerful calculational techniques, usually in the
context of driven vertical transport, basing their work on resonant
states and resonant Wannier-Stark functions of a
system~\cite{Gluck_review, Toshima05}; however, the physical
mechanism underlying Rabi oscillations has not been sufficiently
elucidated.

We avoid many simplifying yet restrictive model conditions, by
directly solving the time-dependent Schr\"odinger equation along
with transparent boundary conditions (TrBC). This enables us to go
beyond the common two-band approximation in the high-field regime,
thereby obtaining a more reliable description of carrier dynamics,
in particular Rabi oscillations, and allowing us to model new
phenomena in quantum transport. This paper is organized as follows.
Section~\ref{sec:model} provides details of the physical model used
and its numerical implementation; the simulation results are
presented in sections~\ref{sec:RO},~\ref{sec:self_interference}, and
\ref{sec:res_123}. Section~\ref{sec:RO} deals with the occurrence
and structure of resonances; section~\ref{sec:self_interference}
discusses their nature, and self-interference of a wavepacket.
Finally section~\ref{sec:res_123} describes the carrier dynamics at a
resonance across three minibands as revealed in our simulations.


\section{\label{sec:model}Physical model}

We consider a layered GaAs/Ga$_{1-x}$Al$_{x}$As heterostructure. The
longitudinal motion of a wavepacket $\Psi(x,t)$ representing a
single electron in a zero-temperature biased superlattice with
potential $V_{SL}(x)$ under constant uniform bias $F\,\equiv\,-eE$
is described by a solution of the time-dependent Schr\"odinger
equation
    \begin{eqnarray}       
    -\frac{\hbar^2}{2mm^*} \nabla^2\,\Psi(x,t) + \Big( V_{SL}(x) +
             xF \Big)\,\Psi(x,t) \ \nonumber\\
    \ = \ E\,\Psi(x,t)
    \label{eq:Schrod}
    \end{eqnarray}
which we solved in the time domain.

Experimentally the excitonic carrier population in the conduction
miniband is created by ultrashort laser pulses~\cite{Shimada04}.
This work deals only with conduction miniband electrons, due to the
fact that holes with their large effective mass are well-localized
and do not demonstrate field-dependent absorption
spectra~\cite{Gluck3}.


\subsection{\label{sec:model:TrBC}Transparent boundary conditions}

Unlike previous studies based on a similar approach~\cite{Bouchard,
Zhao00, Diez98}, we used TrBC~\cite{Arnold01, Arnold03, TBC_other}
that were recently derived for the Schr\"odinger equation in 1D. We
followed Moyer~\cite{Moyer} who used Crank-Nicholson to advance the
time, and the Numerov method for the space dependence. This scheme
has recently been employed by Veenstra et al.~\cite{Veenstra}. TrBC
allow one to limit the size of the space within which the numerical
solution proceeds, without artificial (e.g. rigid-wall) boundary
conditions. In addition, modern computational power enabled us to
perform more robust simulations which revealed aspects that have not
been addressed before.

\begin{figure}                  
        \includegraphics [width=4cm,angle=270]
                   {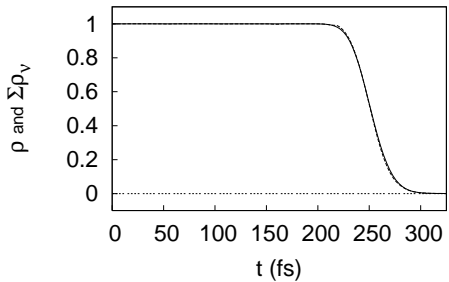}
        \caption[]{Demonstration of transparent boundary conditions. Solid line shows the integrated probability $\rho(t)$
                    remaining inside the computational domain; dashed line shows the sum of occupancy functions $\rho_{\nu}(t)$
                    over the set of basis functions $\{w_{\nu}\}_{\nu=1}^7$ from sample A.}
        \label{fig:TrBC_test}
    \end{figure}

For this work, an extension of the discrete TrBC as described
in~\cite{Moyer} was required to accommodate unequal saturation
potentials on either end of the system (also applicable in case of a
time-dependent potential in the inner region). Details of the finite
element implementation are described in appendix~\ref{appendix:TrBC}.
To demonstrate perfect transmission through the domain border, we
considered a Gaussian wavepacket with initial width of 50 nm, that
was set free to slide down a linear potential ramp with slope
$F=3~\invunits$ (Fig.~\ref{fig:TrBC_test}). As the wavepacket crosses
the border, the total probability remaining in the domain falls
smoothly to zero, showing that there is no reflection from the
perfectly transparent boundary.

\subsection{\label{sec:model:potential}Superlattice potential}

In order to avoid abrupt steps in the potential profile, we replaced
the commonly used square barriers by an analytic form
    \begin{eqnarray}                
    V_{SL}(x) &=& \frac{V_0}{2} \, \Big[ \,  \tanh \frac{x+a/2}{\sigma} \,
- \, \tanh \frac{x-a/2}{\sigma} \, \Big]
    \label{eq:tanh_potl}
    \end{eqnarray}
This is also a more realistic representation of an actual
heterostructure potential~\cite{Gluck3}.

\begin{figure}                      
        \includegraphics [height=2.5cm,angle=270,keepaspectratio=true]{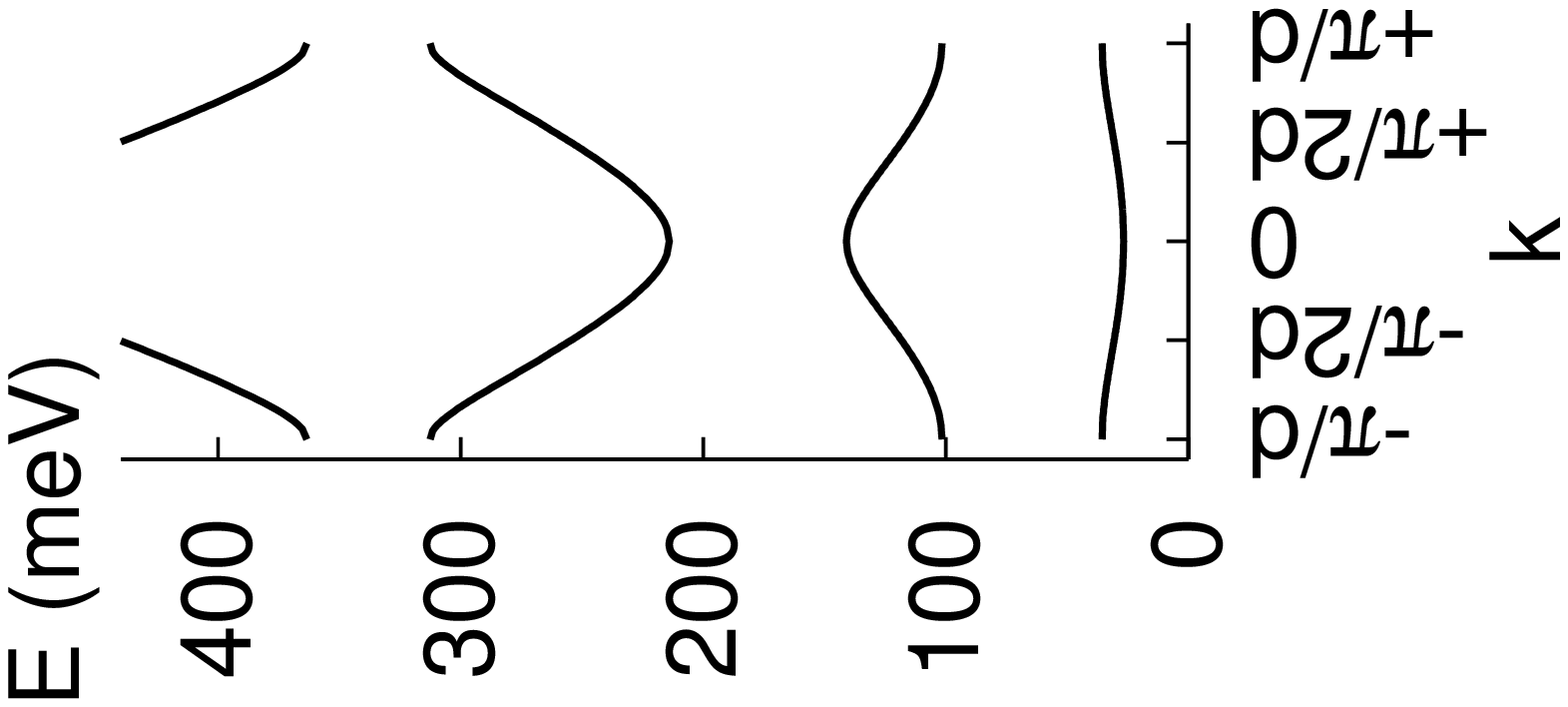}
        \includegraphics [height=2.5cm,angle=270,keepaspectratio=true]{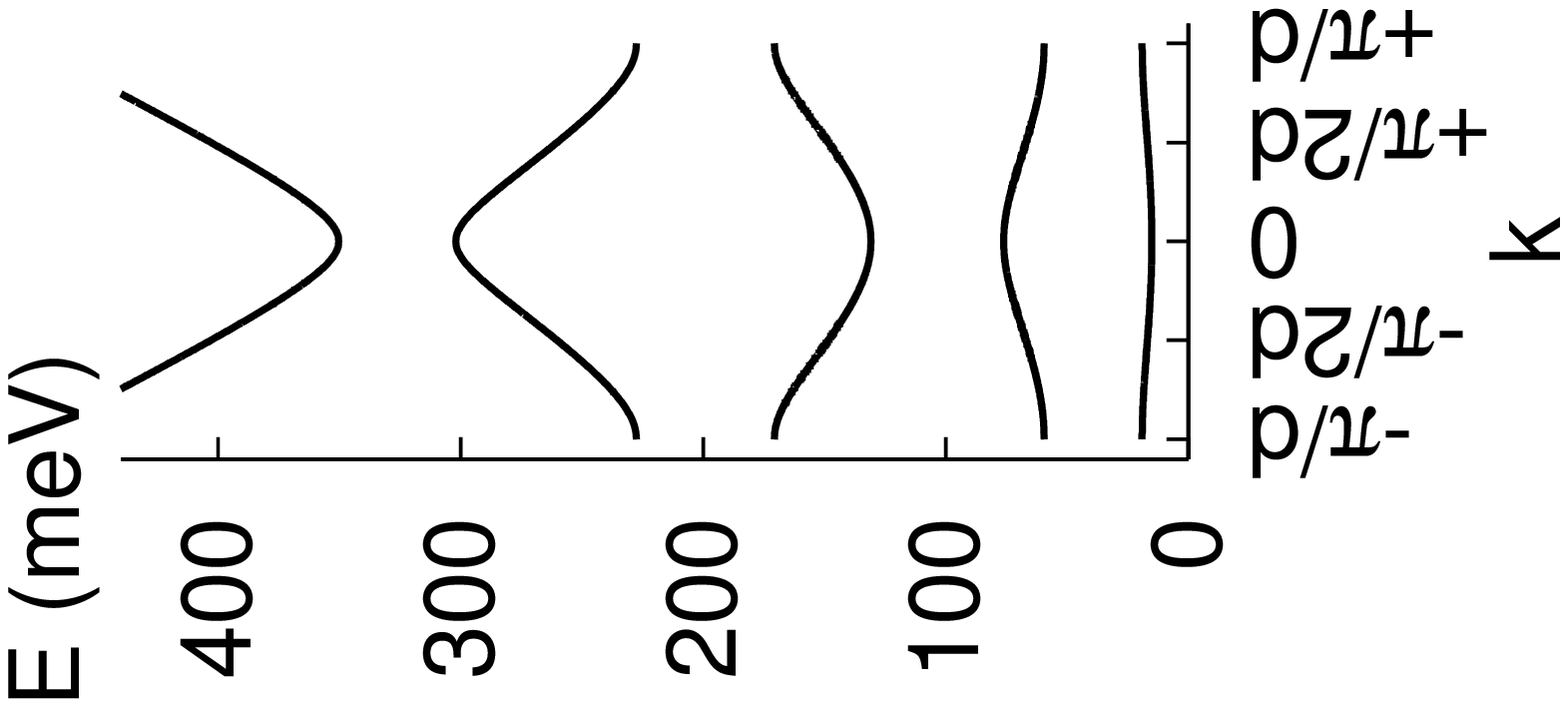}
        \includegraphics [height=2.5cm,angle=270,keepaspectratio=true]{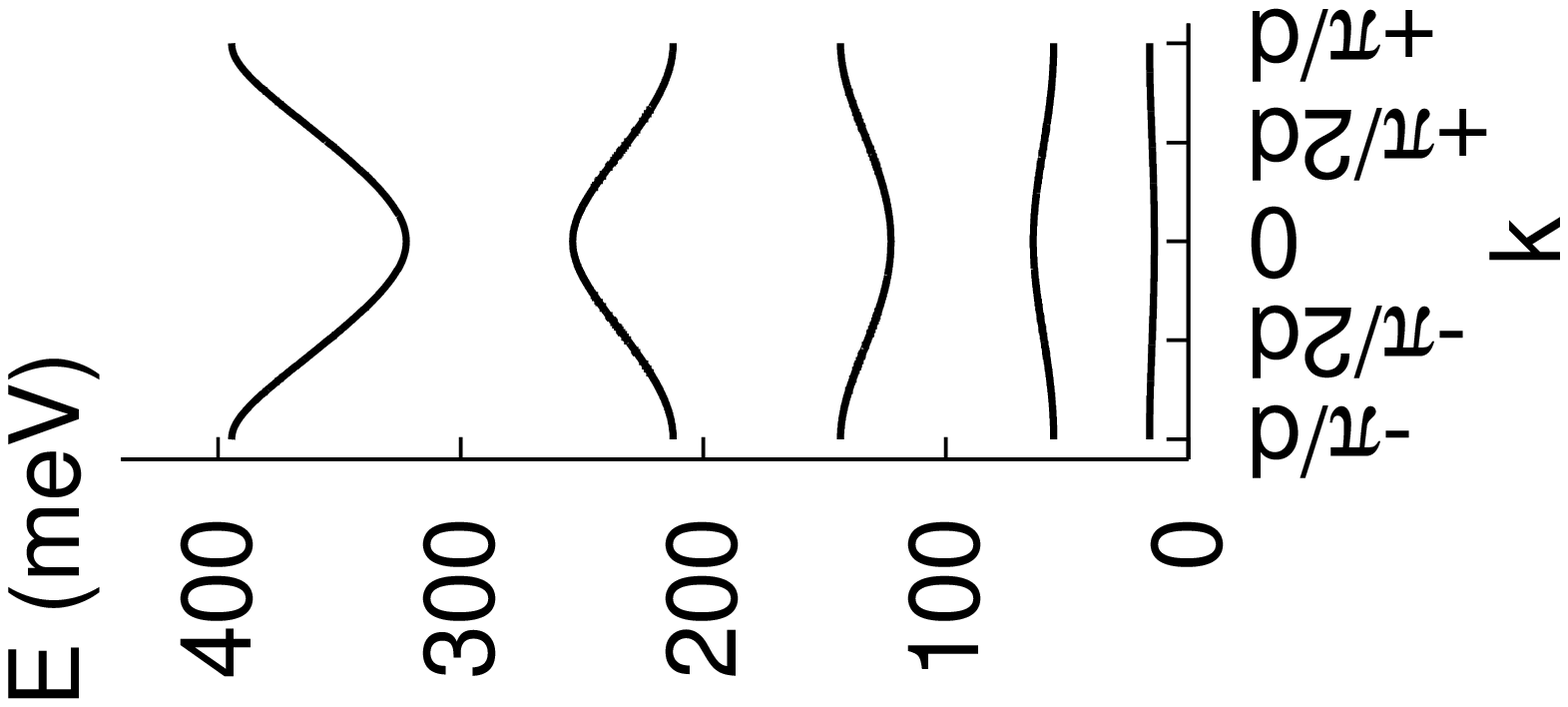}
      \caption[]{Band structure of the superlattice samples studied (from left to right: samples A, B, and C).}
      \label{fig:Bands}
    \end{figure}

We modeled a typical GaAs/Ga$_{1-x}$Al$_{x}$As
heterostructure~\cite{pacher03} in the envelope function
approximation. It has monolayer (ML) thickness of 0.283 nm and
barrier height $V_0=790\,x $ meV, $x$ being the fraction of Al. The
average electron effective mass was set at  $m^*=0.071$, to take
account of non-parabolicity. The system has little sensitivity to the
parameter $\sigma$ over the range $0.2 \to 0.5$ nm. We chose $\sigma$
= 0.4 nm for all our samples, so that 80\% of the potential barrier
height rises over two monolayers. The characteristics of the
potentials used in our simulations are laid out in Fig.~\ref{fig:Bands}
and Table~\ref{tb:tanh_shapes}.

Using TrBC assumes that the potential outside of the considered
region is constant. Our investigations focussed on the dynamics of
the wave packet inside the biased superlattice rather than those of
the emitted part. The length of the computational domain, at least 40
cells, was sufficient for the simulations, since it allowed space for
the packet to move while avoiding any contribution from surface
states. Calculations over a wide range of bias with 40-cell and
120-cell domains showed no appreciable difference (less than
0.1\%) in the output. Although the superlattices considered
are ideal, one could introduce imperfections such as doping and
barrier thickness fluctuations, in order to study their dephasing
effect on electronic coherence.

    \begin{table}
      \begin{center}
            \begin{tabular}{ccccc} \hline \hline
              Sample& $V_0$, meV& $d$, nm(ML) & $a$, nm(ML) & $\sigma$, nm \\ \hline
                A &  212       &  13.0 (46)  &  3.1 (11)   &   0.4        \\
                B &  250       &  17.3 (61)  &  2.5 (9)    &   0.4        \\
                C &  350       &  19.0 (67)  &  2.5 (9)    &   0.4        \\ \hline \hline
            \end{tabular}
      \end{center}
      \caption[]{Geometric parameters of the superlattice samples studied.
                 Barrier height of 100 meV corresponds to $x$=0.13, of 212 meV to $x$=0.18, of 250 meV to $x$=0.3
                 and of 350 meV to $x$=0.44 in the GaAs/Ga$_{1-x}$Al$_{x}$As structure.}
      \label{tb:tanh_shapes}
    \end{table}


\subsection{\label{sec:model:data}Data analysis}

In this work, we will denote a Wannier-Stark (WS) level centered on
the well with index $k$ and belonging to miniband $\nu$ as
$E_{\nu}^k$, the corresponding WS wavefunction being $W_{\nu}^k(x)$
($\nu$ = 1,2,\ldots); the well about which the initial wavepacket is
centered is assigned index 0. By ``resonance" (denoted as
$\R_{\nu\Btw\mu}^n(X)$) we will mean an anticrossing of energy levels
$E_{\nu}^k$ and $E_{\mu}^{k+n}$ belonging to WSL$\nu$ (the $\nu^{th}$ Wannier-Stark ladder, WSL) and
WSL$\mu$, respectively, in sample X, where X can be A, B or C (so
that $E_{\nu}^k\,+\,n\,Fd\,=\,E_{\mu}^{k+n}$, where $\nu,\,
\mu\,=\,1,2,\ldots$ and index $n\,=\,1,2,\ldots$). In terms of bias
values, the term ``resonance" will refer to a range of bias values
that are close to the resonant bias ($F\,=\,F_{\R_{\mu\Btw\nu}^n}$ or
$F_n$ for $\R_{\mu\Btw\nu}^n$) and for which Rabi oscillations
and/or RZT can be resolved. In case some of those indices are of
little importance or have already been specified, they are
omitted for brevity. Throughout this work, the time unit is
conveniently chosen to be the Bloch period at given bias:
$T_B=\frac{2\pi\hbar}{Fd}$ and the length unit is the cell width $d$,
unless stated otherwise.

To visualize the interminiband dynamics of a wavepacket $\Psi(x,t)$,
we define the absolute occupancy functions
    \begin{eqnarray}                
    \rho_{\nu}(t) &=& \sum_k \, |\langle\Psi(x,t)|W_{\nu}^k(x)\rangle|^2,
    \label{eq:occAbs}
    \end{eqnarray}
which is the wavepacket intensity projected onto the $\nu^{th}$
tight-binding (TB) miniband at time $t$ (the index $k$ runs over the
cells within the computational domain) and the relative occupancy
functions $\frac{\rho_{\nu}}{\rho}(t)\,$$\equiv$
$\,\frac{\rho_{\nu}(t)}{|\Psi(x,t)|^2} \in [0,1]$; $\nu=1,2,\ldots$

Owing to the method of their construction, both the TB Wannier-Stark
and Wannier states include the same harmonics, i.e. Bloch
functions~\cite{Rossi88}. The main features of projection on
minibands, such as resonant bias values and the Rabi oscillation
period, using either set, were close to indistinguishable (with
difference not exceeding 1\%) for the range of fields considered.
Thus we adopted the simplification of using Wannier functions
$w_{\nu}^k(x)$ as the projection basis.

Generally speaking, the projection-on-minibands method does not apply
at high fields where Wannier-Stark and miniband transport
models~\cite{Wacker_review02} do not hold any more, and a sequential
tunneling model has to be considered. Also, at the points of WSL
energy level anticrossings, the considered TB WS
functions~\cite{Rossi88} cannot reliably describe resonant states
(see e.g.~\cite{Gluck_review, Toshima05}). However, at any bias the
method gives a picture of the wavepacket's distribution in energy (and
hence between wells in real space), since $w_{\nu}(x)$ contain only
harmonics with wavelengths $\lambda \in [\frac{\nu}{2}d,
\frac{\nu+1}{2}d]$. For simplicity, we will use Wannier functions as
the convenient orthogonal basis for miniband projection.


\section{\label{sec:RO}Rabi oscillations: \protect\\ overview}

    \begin{figure}              
      \leavevmode
        \includegraphics[draft=false, height=4.8cm,angle=0,keepaspectratio=true]{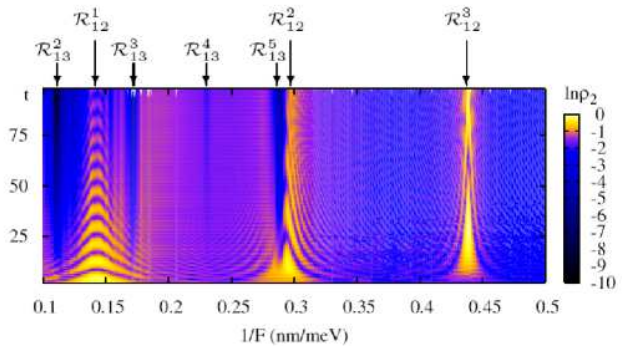}
      \caption[]{(Color online) Absolute occupancy function map for the second miniband in sample A; $\Psi(x,0)=w_1(x)$.}
      \label{fig:Toshima_W1_bands}
    \end{figure}

A typical case of interminiband carrier dynamics is illustrated in
Fig.~\ref{fig:Toshima_W1_bands}. The time evolution of
$\rho_{\nu}(t,F)$ is shown on a map plot comprising results from a
large number of time-dependent simulations over the range of bias
$\frac{1}{F}=0.1\ldots0.5~\units$ with the step
1~$\frac{\textrm{nm}}{\mu \textrm{eV}}$; a greater value of
$\rho(x,t)$ is shown in a lighter color. The large uniformly shaded
areas correspond to exponential Zener decay of the wavepacket out of
the superlattice potential. The darker vertical stripes originate
from resonant Zener tunneling where the wavepacket decays extremely
quickly at energy level anticrossings. In the TB approximation, the
$n^{th}$ anticrossing between WSL$\mu$ and WSL$\nu$ occurs when the
condition $E_{\nu}-E_{\mu}=F_n\,nd\equiv F_1\,d$ is
satisfied~\cite{Shimada_04}. In Fig.~\ref{fig:Toshima_W1_bands},
the period in $\frac{1}{F}$ of the series of stripes marked on top by
shorter arrows implies interminiband separation
$F_1\,d=(223\pm 5)$~meV and clearly corresponds to the anticrossings
associated with $\R_{1\Btw3}$, since $E_3-E_1=227.2$~meV from our TB
calculations.

$\R_{1\Btw3}$ do not show strong  Rabi oscillations, since the
energy gap following the third miniband, $\Delta E_3$, is only 50~meV
while an electron easily overcomes the interminiband separation
$E_2-E_1=89$~meV. Therefore $\Delta E_3$ is too small to strongly
bind an electron in the superlattice. On the other hand, $\Delta
E_2=73.02$~meV suppresses tunneling to the third miniband at lower
biases, and only $\R_{1\Btw3}^{k\leq 5}$ are seen.

There is also a set of prominent periodic spikes marked by longer
arrows, corresponding to the group of resonances
$\R_{1\Btw2}^{1\ldots3}$. In contrast to $\R_{1\Btw3}$, the
resonances $\R_{1\Btw2}$ do exhibit oscillations in $\rho_2(t)$ with
time, corresponding to interminiband Rabi oscillations; they are
wider and demonstrate a higher RZT rate for lower indices. The
values $F_{\R_{1\Btw2}^2}$=(6.9$\pm$0.2)~$\invunits$ and
$F_{\R_{1\Btw2}^3}$=(2.33$\pm$0.05)~$\invunits$ are reasonably close
to the anticrossing calculated in~\cite{Toshima05}
(7.2~$\invunits$ and 2.4~$\invunits$, respectively), even though the
potential considered here was not exactly the square barrier one
used in that work.

Minor periodic changes in $\rho_2(t)$, creating a light horizontal
mesh on the background with period $T_B$, are the signature of Bloch
oscillations. For extremely high fields ($F>10~\invunits$) the period
and magnitude of these oscillations explodes (see e.g. the left edge
of Fig.~\ref{fig:tanhHigh_W1_band1}), since $\Delta E_1 < F(d-a)$ and
a transition to the next higher miniband can be made without
tunneling; then a theory different from interwell hopping must be
applied.

Generally, it was found that energy anticrossings do not necessarily
result in strong RZT for a strongly bound interacting WSL. Rabi
oscillations appear to be the reverse side of RZT in the WSL
interaction: for strong RZT, overdamped Rabi oscillations are seen
(i.e. $\R_{1\Btw3}$ in the above example), while persistent and strong
Rabi oscillations correspond to weak RZT (e.g. $\R_{1\Btw2}$). A
quantitative relation between the two depending on the strength of
the potential and resonance index remains an open question; up to now
Rabi oscillations have been studied separately from RZT.


\subsection{\label{sec:RO:shape}Resonance shapes}

\begin{figure}[b]                   
        \begin{picture}(8,8)(0,0)
          \put(-0.5,2.5){  \includegraphics[height=8cm,angle=270,keepaspectratio=true]{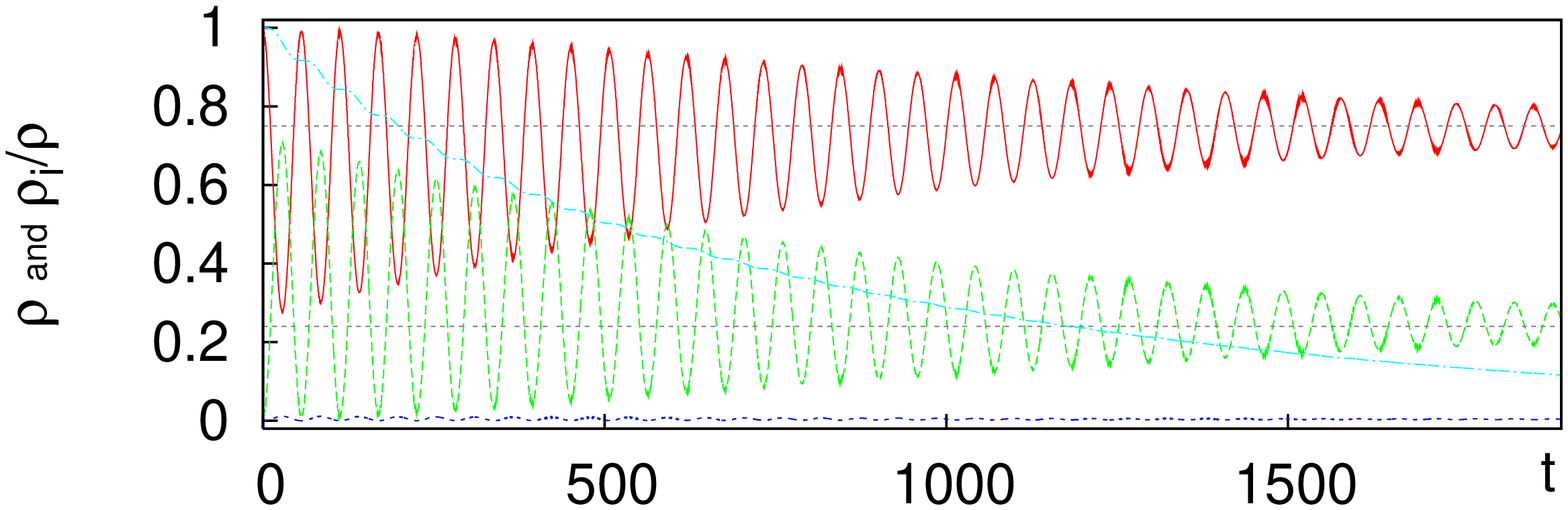} }
          \put(0,2.75){    \includegraphics[draft=false,height=5cm,width=4cm,angle=0,keepaspectratio=false]{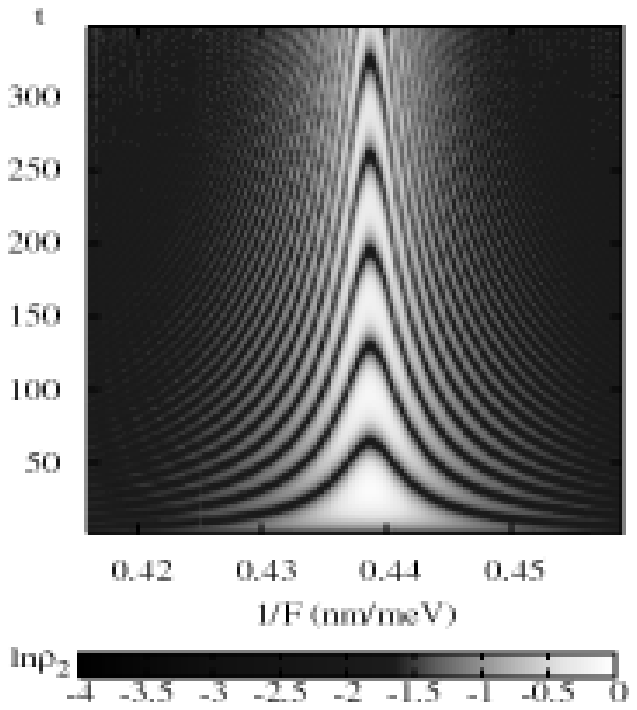}}
          \put(4.5,3){     \includegraphics[width=3.3cm,angle=0,keepaspectratio=true]{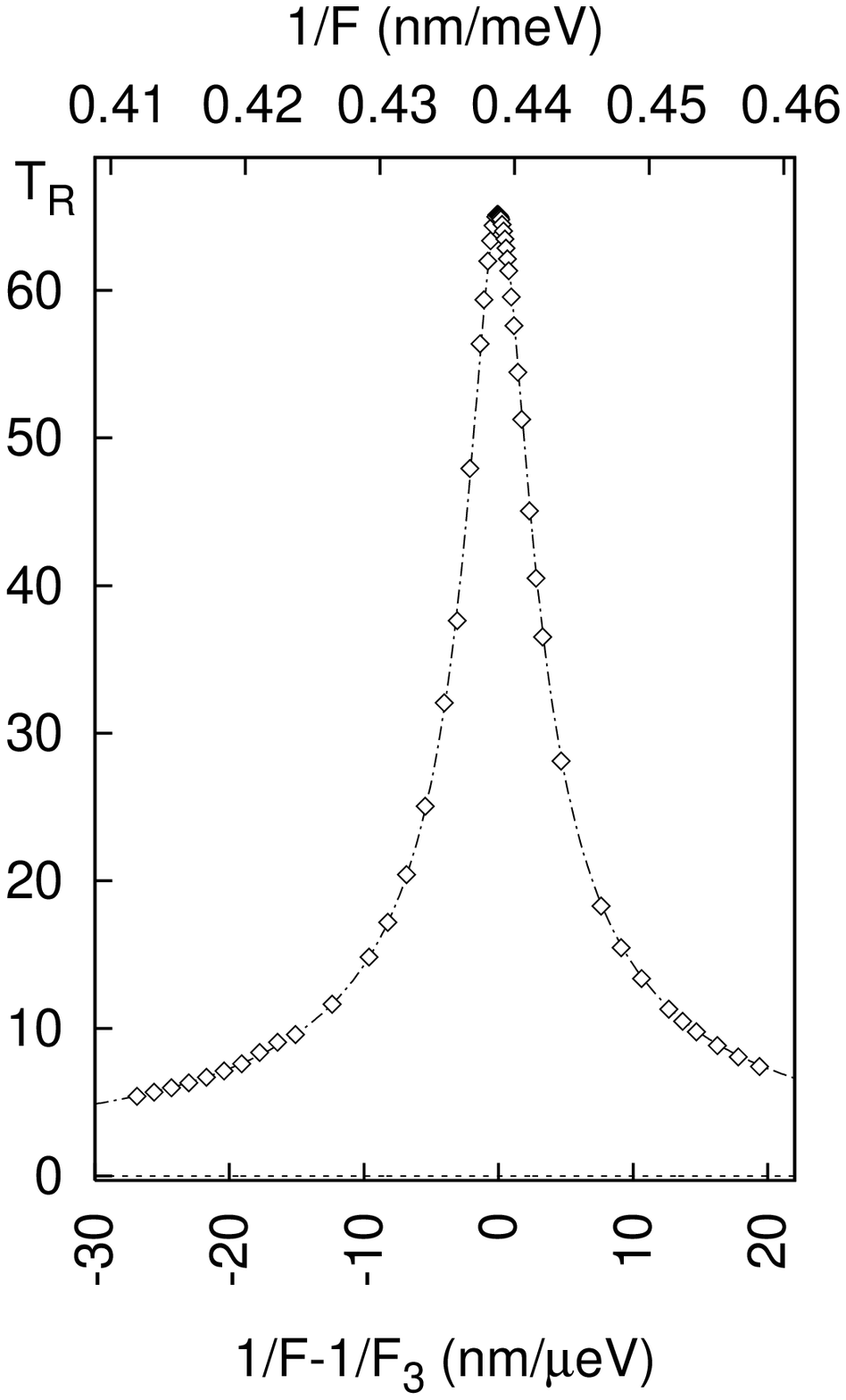}}
        \end{picture}
      \caption[]{(Color online) Detailed view of $\R_{1\Btw2}^3$ in sample A (top left). Period of Rabi oscillations vs. bias for $\R_{1\Btw2}^3(A)$ is fitted by means of a Lorentzian as in Eq.~\protect\ref{eq:T_fit} (top right).
                  Relative occupancy functions and wavepacket norm
                  are shown for the bias $F$ such that $|\frac{1}{F}-\frac{1}{F_{\R_{1\Btw2}^3}}|=1.5\
                  \Gamma_{\R_{1\Btw2}^3}$; $\rho_1$ is shown in solid (red), $\rho_2$ in dashed (green), $\rho_3$ in dotted
                  (blue), $\rho$ in chain-dotted (magenta) lines, and asymptotic occupancy values in horizontal lines (bottom).}
      \label{fig:res3_details}
    \end{figure}

The detailed structure of a resonance is shown in
Fig.~\ref{fig:res3_details} which is an enlargement of the resonance
$\R_{1\Btw2}^3(A)$ from Fig.~\ref{fig:Toshima_W1_bands}. Near a
resonance, one sees Rabi oscillations as persistent oscillations of
significant magnitude in $\rho_2(t)$ with period $T_{\R}\sim \, 10
\ldots 100\, T_B$ (the corresponding frequency range is between the
microwave and infrared regions). Both vertical and horizontal
cross-sections of the plot demonstrate periodic oscillations of
$\rho_2$.

A vertical cross-section of $\R_{1\Btw2}^3(A)$ at a near-resonant
bias, shown at the bottom of Fig.~\ref{fig:res3_details}, explicitly
demonstrates Rabi oscillations. Oscillations in $\rho_1(t)$ and
$\rho_2(t)$ are shifted in phase by $\pi$, which
means that their coupling shows up as Rabi oscillations rather than
RZT. In the example $\R_{1\Btw2}^3(A)$, the third miniband
contributes little to total wavepacket norm. Its population could be
due to two factors: (i) the non-TB WS functions of the second
miniband involve harmonics from the third TB miniband, and (ii) the
escaping part of the wavepacket passing through the third miniband
on its way to the continuum. Since $\rho_3(t)$ oscillates in phase
with $\rho_2(t)$, we conclude that the second factor is dominant in
this example.

When the asymptotic values of $\rho_1(t)$ and $\rho_2(t)$ in the
limit $t\rightarrow\infty$ are not equal, as in the bottom panel of
Fig.~\ref{fig:res3_details}, we are dealing with a so-called
asymmetric decay~\cite{Fidio00}: the minibands 1 and 2 are coupled
to the continuum to a different extent. Right at a resonance bias value,
the extremely strong interaction between the two minibands
essentially merges them, and their asymptotic population values were
both very close to ${1}/{2}$. Away from the resonant
bias, the coupling is weaker and the difference between
the two values increases, eventually approaching unity.

\begin{figure}[t]                   
      \includegraphics[draft=false,height=5.5cm,angle=0,keepaspectratio=true]{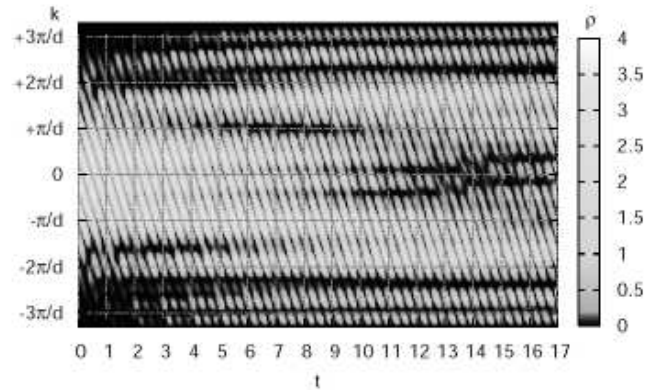}
      \caption[]{(Color online) Dynamics of the wavepacket having initial form $\Psi(x,0)=w_1(x)+w_1(x+3d)$ in reciprocal space at $\R_{1\Btw2}^2(A)$.}
      \label{fig:psi_tunneling_2W1}
    \end{figure}

The dynamics of $\Psi(x,t)$ in k-space demonstrates features similar
to those of Bloch oscillations across a single miniband, i.e. most
of the wavepacket steadily traverses the first and the second
minibands as a whole with period $2\,T_B$ as shown in
Fig.~\ref{fig:psi_tunneling_2W1}. The map plot consists of a series
of images of the total probability density $\rho(x,t)$. We chose
$\Psi(x,0)$ to be a linear combination of two Wannier functions in
order to reduce the wavepacket uncertainty in k-space and to get a
better resolution of fine details. From this perspective, an
anticrossing can be thought of as a phenomenon of two adjacent WSL
merging into a single broader one.

For all resonance indices and samples examined, the Rabi oscillation
period $T_{\R}$ of an $n^{th}$ resonance clearly showed a
Lorentzian-like dependence as anticipated from the nature of Rabi
oscillations~\cite{Akulin}:
    \begin{eqnarray}                
      T_n(F) &=& T_n^{max} \, \Bigg[\bigg(\frac{\frac{1}{F}-\frac{1}{F_{\R_{\nu\Btw\mu}^n}}}{\Gamma_{\R_{\nu\Btw\mu}^n}}\bigg)^2 + 1\Bigg]^{-\frac{1}{2}}
      \label{eq:T_fit}
    \end{eqnarray}
which is shown in the top right panel of Fig.~\ref{fig:res3_details}.
The data were obtained from a fit of $\rho_2(t)$ at
$\R_{1\Btw2}^3(A)$ over a long length of time considered
($t=50\ldots100~T_{\R}$). When speaking of half-width at half-maximum
(HWHM, or $\Gamma_{\R_{\mu\Btw\nu}^n}$), we will be referring to
HWHM of the corresponding fit to $T_n(F)$ at $\R_{\mu\Btw\nu}^n$.

\begin{figure}[t]                   
        \includegraphics[height=4cm,angle=0,keepaspectratio=true]{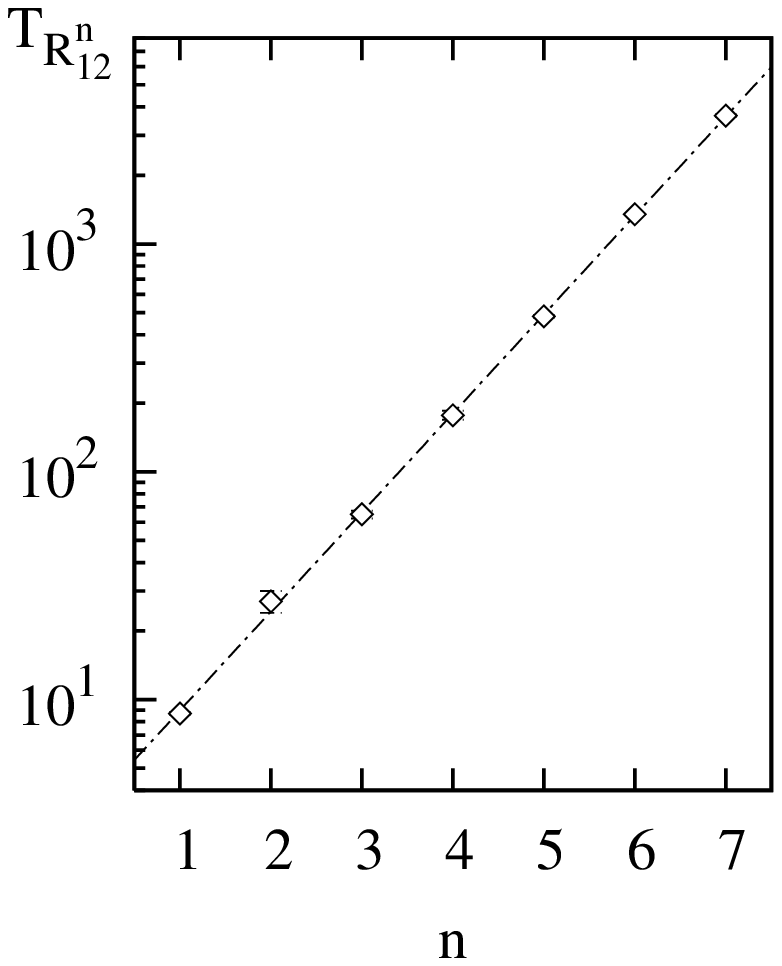}
        \includegraphics[height=3.9cm,angle=0,keepaspectratio=true]{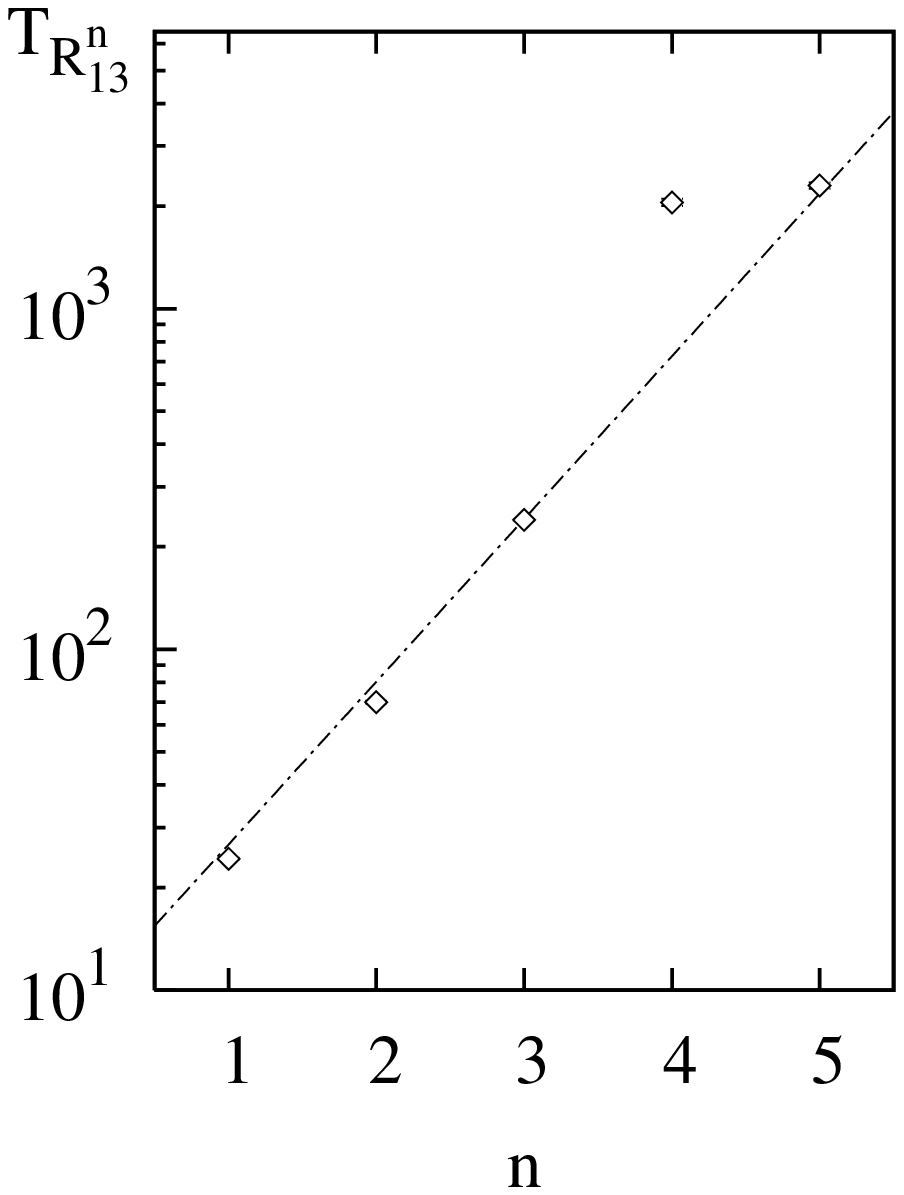}
        \caption[]{Logarithmic fit of $T_n^{max}$ versus resonance index $n$ for $\R_{1\Btw2}(A)$ (left) and $\R_{1\Btw3}^n(C)$ (right).}
        \label{fig:T_index_fit}
    \end{figure}

The maximum Rabi oscillation period observed at a resonant bias,
$T_n^{max}$, was found to grow exponentially with resonance index $n$
in the following way:
    \begin{eqnarray}                
      T_n^{max} &=& T_1^{max} \bigg( \frac{T_2^{max}}{T_1^{max}} \bigg)^{(n-1)}, \;\;\; n=1,2,\ldots
    \label{eq:fit_T_n}
    \end{eqnarray}
(see left part of Fig.~\ref{fig:T_index_fit}), which was
expected~\cite{Gluck5} since the transition matrix element $|\langle
W_1^k(x)|W_2^m(x)\rangle|^2\,~\propto~e^{-|k-m|}$. As the bias is
reduced, the horizontal tunneling channel to the next WSL lengthens,
and it takes longer for the probability density to build up in the
other miniband. We remark that perturbation theory predicts the
dependence in Eq.~(\ref{eq:fit_T_n}) to be linear~\cite{Diez98},
but that does not apply at the high bias considered.

\subsection{\label{sec:RO:pulsed}Pulsed output from the system}

\begin{figure}[b]               
      \leavevmode
          \includegraphics[height=6cm,angle=270,keepaspectratio=true]{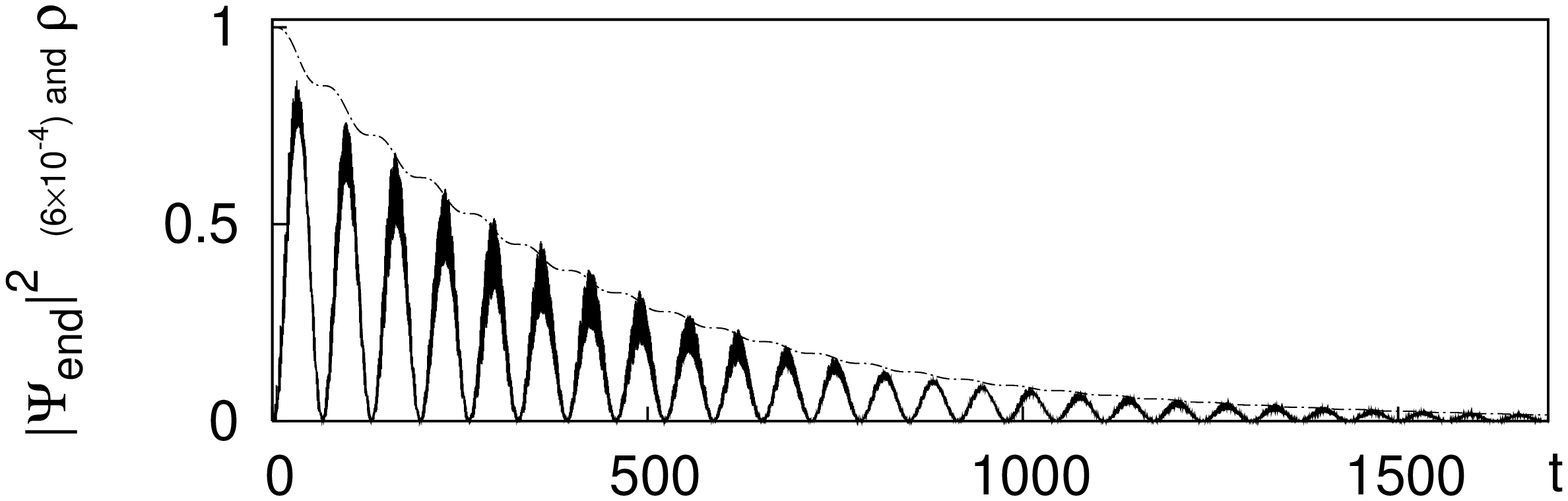}
          \includegraphics[height=6cm,angle=270,keepaspectratio=true]{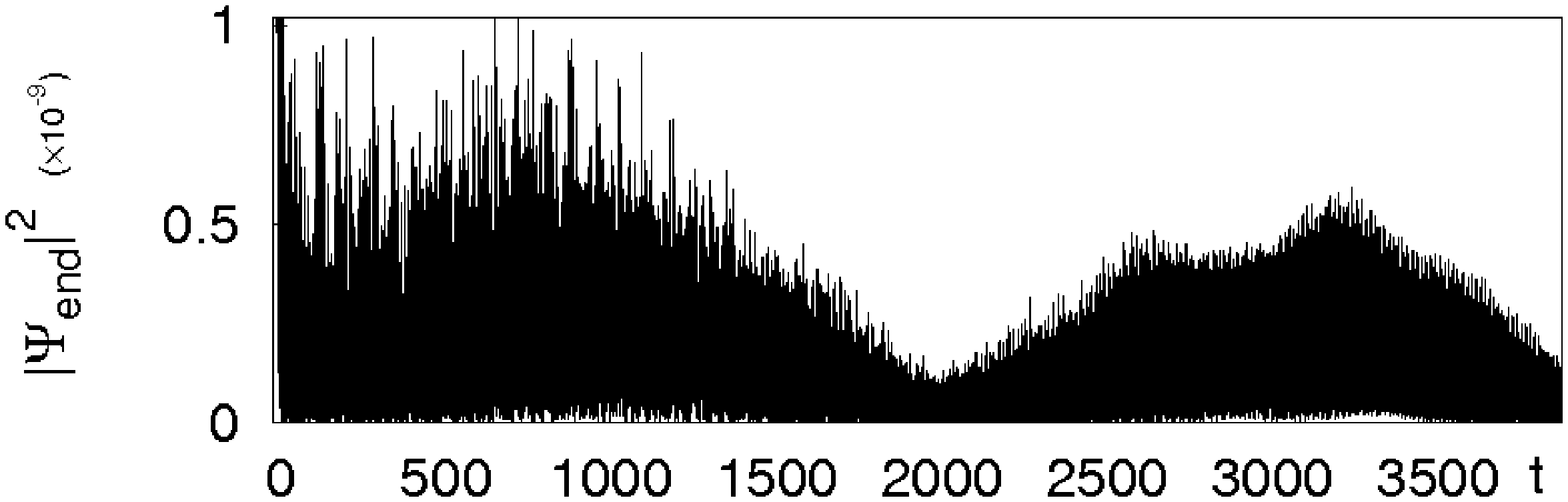}
      \caption[]{Pulsed output of a Rabi oscillating system at $\R_{1\Btw2}^3(A)$ (top) and $\R_{1\Btw3}^4(C)$ (bottom);
                chain-dotted line is the wavepacket norm.}
      \label{fig:Psi_End_Rabi}
    \end{figure}

Rabi oscillations of the carrier produce a periodic coherent pulsed
output with period $T_{\R}$ similar to that of Bloch
oscillations~\cite{Gluck_review}. The data shown in
Fig.~\ref{fig:Psi_End_Rabi} correspond to a record of
$|\Psi_{end}|^2~\equiv~|\Psi(x_{end},t)|^2$ values over time, at the
endpoint $x=x_{end}$ of the superlattice having a lower saturation
potential.

In the case of a relatively short Rabi oscillation period, the
output consists of damped sinusoidal oscillations decaying at the
same rate as the wavepacket norm. For a long Rabi oscillation
period accompanied by weak RZT, as in the case of
$\R_{1\Btw3}^4(C)$, the pulsed output is significantly distorted.
Over the course of a single oscillation, the net effect of
relatively small factors becomes significant; in the bottom section
of Fig.~\ref{fig:Psi_End_Rabi}, Rabi oscillations cause the drops in
$|\Psi_{end}|^2$ with period $T_{\R}=2050~T_B=186~ps$. Away from a
resonance, the output pulses flatten out and on, a time scale longer
than $T_B$, only a smooth exponential decay in accordance with Zener
theory would be observed.

The form of the pulsed output is determined by the strength of Rabi
oscillations, by their rate of magnitude decay, and by the strength of
RZT. In principle, one could experimentally observe Rabi
oscillations as well as establish the relation between Rabi
oscillations and RZT, by measuring the system's pulsed output.


\section{\label{sec:self_interference}Wavepacket self-interference}

A detailed consideration of the interference mechanism allows us to
analyze resonance phenomena and draw some important conclusions
without the need to calculate the system's spectrum.

A strong correlation between subsequent well-to-well tunneling
events arises due to their coherence. This correlation manifests
itself in the interference between Bloch oscillations and intrawell
oscillations (first resolved in~\cite{Bouchard}) with frequencies
$\omega_B$ and $\omega_{\mu\Btw\nu}$ that is constructive for
near-resonant bias and destructive otherwise, with frequency
detuning per single intrawell oscillation cycle
    \begin{eqnarray}                
      \Delta\omega \,=\,\omega_{\mu\Btw\nu}-n\omega_B &=& \frac{(E_{\mu}-E_{\nu})-Fnd}{\hbar}
      \label{eq:delta_w}
    \end{eqnarray}
around a resonance with index $n$. If constructive, the interference
leads to a gradual transition of the center of mass of probability
density between the 0$^{th}$ and the $n^{th}$ cells and between the
$\mu^{th}$ and the $\nu^{th}$ Brillouin zones in k-space.

    \begin{figure}[t]               
          \includegraphics[height=7cm,angle=270,keepaspectratio=true]{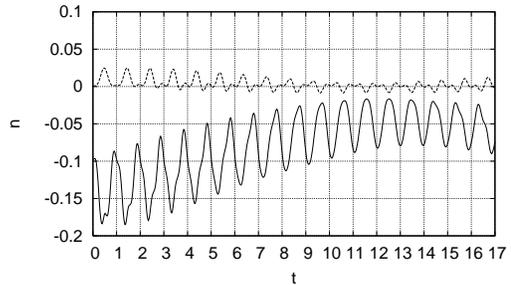}
      \caption[]{Dynamics of the electron dipole moment $\langle\Psi(x,t)|x|\Psi(x,t)\rangle$ calculated over the 0$^{th}$ (solid line) and the second (dashed line)
           cells at $\R_{1\Btw2}^2(A)$. the wavepacket $\Psi(x,0)=w_1(x)$ was taken at a near-resonant bias so that $\frac{T_{\R}}{2}=13.5~T_B$ ($\frac{1}{F}=0.296~\units$,
            $\frac{1}{F}-\frac{1}{F_{\R_{1\Btw2}^2}}$ = 0.8 $\Gamma_{\R_{1\Btw2}^2}$); position $n$ is taken with respect to the center of the
            corresponding cell, in units of the cell width.}
      \label{fig:psi_dephasing_IO}
    \end{figure}

At a resonant bias, the saturation value of occupancy functions of
the two coupled minibands approached unity. Off resonant bias, the
transition of probability density from one miniband to another was
observed to reverse direction when a dephasing of $\pi$ had
accumulated between the two resultant oscillations of wavepacket
components residing in the two coupled minibands
(Fig.~\ref{fig:psi_dephasing_IO}). One can clearly see intrawell
oscillations (especially around $t=0$) and Bloch oscillations, as
well as the process of dephasing between Bloch and intrawell
oscillations. If we call the wavepacket piece residing in the
$n^{th}$ cell $\Psi_n$, then with time, the phase shift between the
peaks of the resultant oscillatory motion of $\Psi_0$ and $\Psi_2$
builds up, and reaches a net increment of $\pi$ at
$t=\frac{T_{\R}}{2}=13.5~T_B$. Remarkably, this interval corresponds
to a maximum of $|\Psi_2(t)|^2$ and hence to a Rabi oscillation
peak.

Given these results, we conclude that interminiband Rabi
oscillations are governed by the process of self-interference of a
wavepacket subject to two intrinsic frequencies: $\omega_B$ and
$\omega_{\mu\Btw\nu}$, and the amplitude of Rabi oscillation at a
given bias is determined by bias detuning from its resonant value.
This again illustrates the coherence of multiwell tunneling, and in
principle one can judge the coherence length of a superlattice by
the number of observable resonances.


\subsection{\label{sec:built_up_state}Built-up state}

    \begin{figure}[b]                   
          \includegraphics[height=4cm,angle=270,keepaspectratio=true]{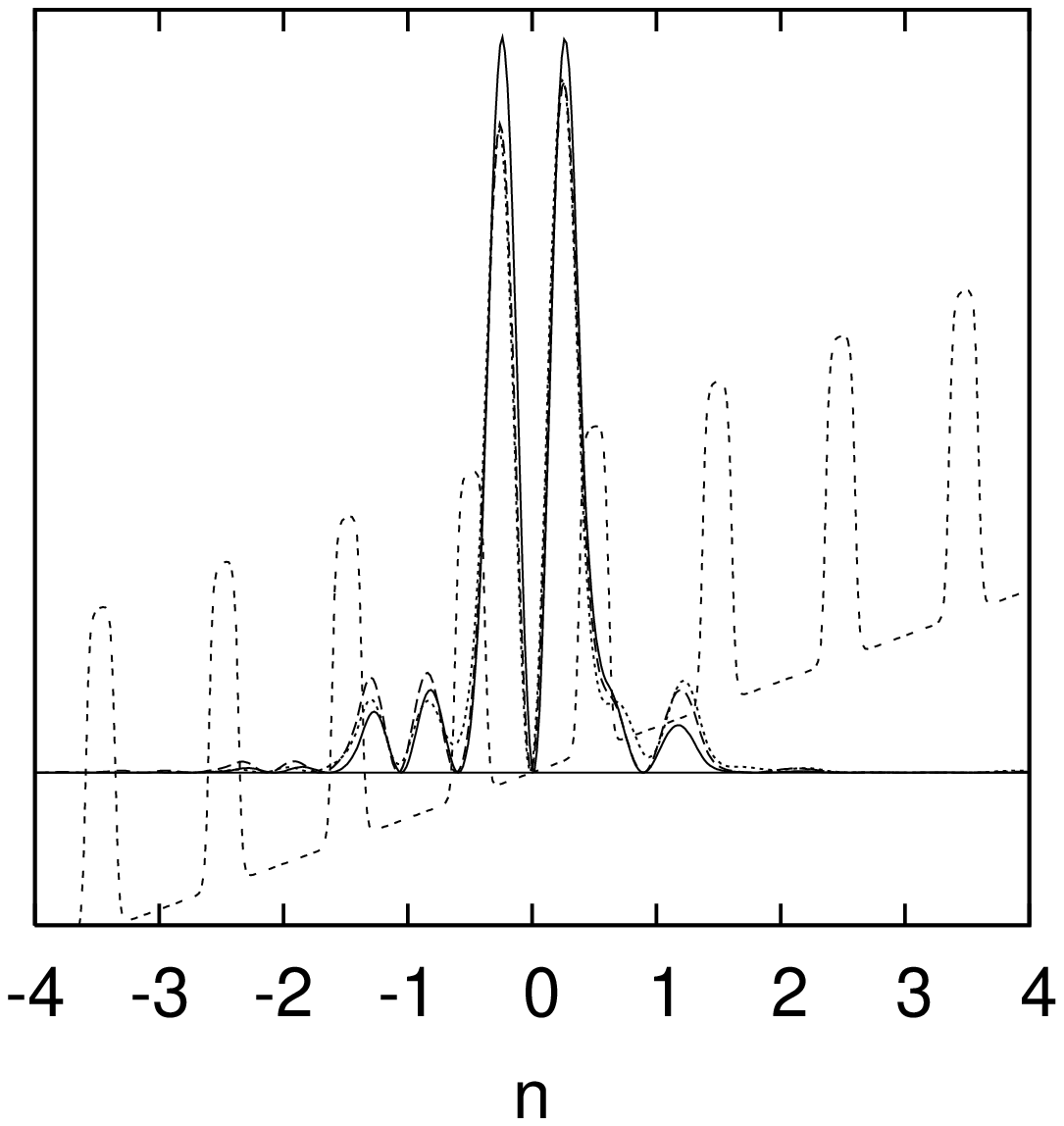}
          \includegraphics[height=4cm,angle=270,keepaspectratio=true]{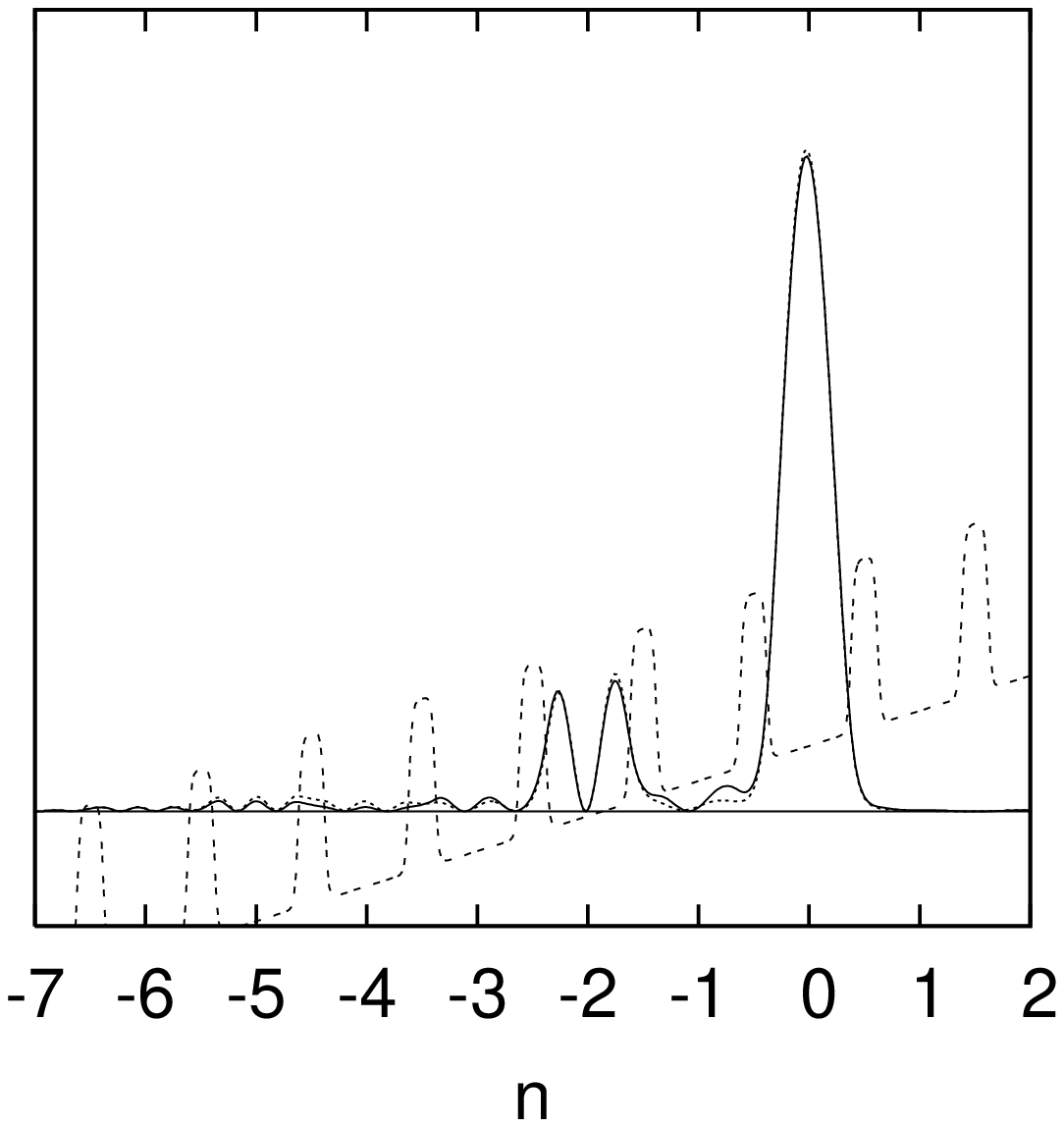}
      \caption[]{Comparison of norms of a built-up state at $F_{\R_{1\Btw2}^2(A)}$ at resonant field for
                 $\Psi(x,0)=w_1(x)$ (dashed line) and $\Psi(x,0)=W_1(x)$ (dotted line)
                 with the corresponding Wannier-Stark state at $\R_{1\Btw2}^3(A)$ (solid line line) in the left panel.
                 In the right panel, the built-up state at a near-resonant bias is in dotted line. For reference,
                 the biased superlattice potential is shown in double-dotted line.}
      \label{fig:built_up_psi}
    \end{figure}

Let us consider the shape of the probability density that is
building up, away from the wavepacket's original location, in the process
of Rabi oscillation (built-up state). Without loss of generality, we
will consider dynamics of $\Psi(x,0)=w_1(x)$ at $\R_{1\Btw2}(A)$ and
will denote by $\Psi_{\R}(x)$ a built-up state in the corresponding
cell $n$. The shape of the probability density $|\Psi_{\R}(x)|^2$,
turns out to be nearly the same for different values of the
off-resonance field, and it also closely resembles the
corresponding TB WS wavefunction, at least when $|\Psi_{\R}(x)|^2$ is at its peak.

The latter comparison is made in the left panel of
Fig.~\ref{fig:built_up_psi}. Although in~\cite{Toshima05} the
corresponding computed resonant wavefunction had significant
presence in the wells with indices $n=-3\ldots-7$, our
$|\Psi_{\R}(x)|^2$ is not large there. A possible explanation is
that the built-up state in the second miniband may include
components from the third miniband as well. $\Psi_{\R}(x)$ appears
to be only slightly more widespread than the TB $W_2(x)$, and on a
linear plotting scale they are quite close as expected
from~\cite{Gluck_review} for the moderate field considered. Note
also that $|\Psi_{\R}(x)|^2$ shows slight asymmetry in k-space, in
agreement with~\cite{Gluck_review}: for example, in
Fig.~\ref{fig:psi_tunneling_2W1} at $t=\frac{T_{\R}}{2}=13.5~T_B$,
there is more probability density present at negative values of $k$
than at positive ones. This tells us that at moderate fields there
is a great similarity between true and TB WS states. For comparison,
the results for $\Psi(x,0)=W_1(x)$ are presented as well. Despite
dissimilar shapes of the initial forms $W_1(x)$ and $w_1(x)$, the
built-up state looks almost the same for both.

The built-up states at resonant and off-resonant values of bias at
$\R_{1\Btw2}^2$ are compared in the right panel of
Fig.~\ref{fig:built_up_psi}. The first one was taken at $F$ such
that the saturation population of the second miniband was
$\rho_2=0.77$ ($\frac{1}{F}-\frac{1}{F_2}=0.5~\Gamma_2$). The second
built-up state was taken at $F=F_2$ at the moment when the
population of the second miniband reached the same value
$\rho_2=0.77$. Their close resemblance reveals that over the range
of near-resonant fields, the build-up mechanism of
$|\Psi_{\R}(x)|^2$ is the same and produces a (rescaled) WS state of
the second miniband at a resonant bias.


\subsection{\label{sec:res_condition}Resonance condition}

In the process of Bloch oscillatory motion, the wavepacket tunnels
out whenever it approaches the end of the Bloch oscillatory domain,
producing a leaking out pulse. As a pulse propagates in space
upon its escape, it scatters on the potential barriers, and some
fraction of a pulse stays trapped in cells outside of the initial
one. If the oscillations of the trapped part of a pulse happen to be
in phase with those of a subsequent incoming pulse, the conditions
for constructive interference are met and the probability density
$|\Psi_{\R}(x)|^2$ builds up in this well.

For realistic fields, $\omega_B<\omega_{\mu\Btw\nu}$ (or
$T_B>T_{\mu\nu}$), and in order for oscillations of two subsequent
pulses to be in phase in the $n^{th}$ well, the equality
    \begin{eqnarray}                
      n &=& \frac{T_B}{T_{\mu\nu}}\,=\,\frac{\omega_{\mu\Btw\nu}}{\omega_B}
      \label{eq:res_condition}
    \end{eqnarray}
must be satisfied: this is the condition for a resonance with index
$n$. A gradual decrease in bias makes $\omega_B$ smaller, thus
increasing $n$ and shifting the location of the built-up state
further down the potential ramp. Provided that
$\hbar\omega_{\mu\Btw\nu}~\equiv~E_{\nu}-E_{\mu}$ and
$\hbar\omega_B~\equiv~F_n\,d=\frac{F_1}{n}d$, for a resonance to
occur between the $\nu^{th}$ and the $\mu^{th}$ minibands,
$n=\frac{E_{\mu}-E_{\nu}}{F_nd}$. In the TB approximation ($E_{\mu}$
and $E_{\nu}$ are independent of $F$), that leads to a well-known
result $F_n\,\propto\,\frac{1}{n}$~\cite{Shimada_04}. This relation
is satisfied surprisingly well as indicated by results in the
following subsection.

From the above argument, it follows that the role of the interband
jumping mechanism, as proposed in~\cite{Fidio00} for Rabi oscillation
dephasing, must be minimal. Indeed, once tunneling of the carrier
into a certain well (and reflection back which doubles the phase
increment of the $\psi_n(x,t)$) puts it out of phase with the rest
of the system, such well will not be largely populated due to
destructive interference. Mathematically, the discrete bias values allowing constructive
self-interference correspond to existence of poles of the system
scattering matrix, which can be employed to successfully
build a WS state for a multiband system~\cite{Gluck1}.


\subsection{\label{sec:res_bias_values}Resonant bias values}

    \begin{figure}[t]                   
      \leavevmode
      \begin{center}
            \includegraphics[height=6cm,angle=0,keepaspectratio=true]{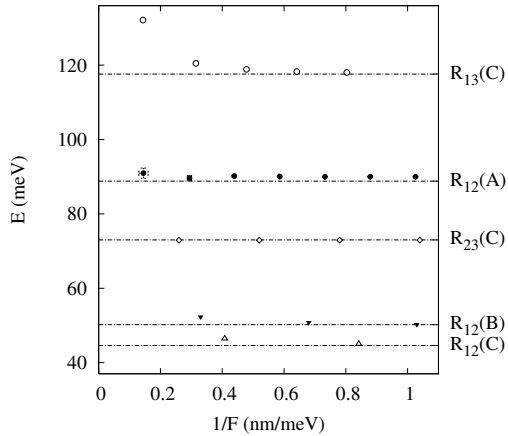}
      \end{center}
      \caption[]{Values of $F_{\R_{\mu\Btw\nu}^n}nd\ $ plotted versus inverse bias for various resonances with lowest indices
          ($n\,=\,1,2,3,\ldots$ ascending from the left to the right) and compared
          with interminiband separation $E_{\mu}-E_{\nu}$ calculated from the tight-binding model.
           Filled circles mark the data corresponding to $\R_{1\Btw2}(A)$,
           filled triangles to $\R_{1\Btw2}(C)$, empty circles to $\R_{1\Btw3}(C)$, empty rhombi
           to $\R_{2\Btw3}(C)$, empty triangles to $\R_{1\Btw2}(B)$.}
      \label{fig:Compare_energy_levels}
    \end{figure}

In order to check the validity of TB model calculations for
predicting resonant bias values at high fields, in
Fig.~\ref{fig:Compare_energy_levels} the values of $nd\,F_n$
obtained from our simulations for the first few resonances in
different samples are compared with the corresponding interminiband
separations calculated in the TB approximation, using a Kronig-Penney
model. The data are plotted versus $\frac{1}{F}$; $x$-error bars for
all points refer to HWHM of a given resonance $\Gamma_n$, whereas
$y$-error bars show broadening of the energy levels and equal
$\Gamma_n$ rescaled in the same manner as $\frac{1}{F_n}$, i.e.
$nd\,\Gamma_n$.

Excluding $\R_{1\Btw3}(C)$, the difference between the \emph{de
facto} relative position of the energy levels found as
$F_{\R_{\mu\Btw\nu}^n}nd$ and that from the TB calculations
($E_{\mu}-E_{\nu}$), is less than 2 meV or $\leq$~5\%, for the
resonances depicted. This is true even for resonances between non-ground
minibands $\R_{2\Btw3}^{k<4}(C)$ (this type of resonance has
experimentally been observed in~\cite{Fidio00}). Thus even at high
fields the equality $F_{\R_{\mu\Btw\nu}^n}nd=E_{\nu}-E_{\mu}$ holds
reasonably well, and the resolved resonances occur periodically in
$\frac{1}{F}$. However, in many cases this difference is seen to be
much larger than the corresponding HWHM which cannot be explained by
broadening of the transition line alone. Because
$F_{\R_{\mu\Btw\nu}^n}nd$ is always larger than
$E_{\mu}\,-\,E_{\nu}$ and this difference increases with bias, it must be
a result of stronger coupling to the continuum at higher bias.
This effect is similar to the energy level structure of a potential
well becoming sparce as the well becomes shallower.

Unlike other resonances, for the series $\R_{1\Btw3}(C)$ the value
of $nd\,F_n$ departs significantly from $E_3-E_2$ for
$\frac{1}{F}\lesssim0.4~\units$. It has been verified that this
deviation is not due to extreme narrowness of the first miniband in
sample C. As follows from the series $\R_{2\Btw3}(C)$, the mutual
alignment of levels belonging to WSL2 and WSL3 is almost unchanged
up to $\frac{1}{F}\gtrsim0.25$; so it is the change in mutual
arrangement of WSL1 and WSL2 that is driving this deviation.

The threshold value $\frac{1}{F}\approx0.4~\units$ corresponds to
the potential drop per cell $Fd\approx47.4$~meV which is close to the
interminiband separation $E_2-E_1=44.6$~meV. According to the TB
model, at $\frac{1}{F}\lesssim0.4~\units$ the energy levels $E_1^0$
and $E_2^1$ would be found in the same cell. This is clearly
impossible due to the Pauli exclusion principle, hence, the
structure of separate Wannier-Stark ladders is completely destroyed
at this point. At such bias, the $w_{\nu}(x)$ correspond only to the set
of Fourier components with certain wavelengths rather than
wavefunctions belonging to certain minibands.

Disruption of the WSL structure seems not to affect
$\R_{2\Btw3}(C)$, however: the initial wavepacket $\Psi(x,0)=w_2(x)$
having only components with wavelength $\lambda\in[\frac{d}{2},d]$
was noticed not to gain any significant components with
$\lambda\in[d,2d]$ over time in our simulations, because of
conservation of energy. In general, the structure and properties of
$\R_{2\Btw3}$ were found to be similar to those of $\R_{1\Btw2}$, at
any bias.


\section{\label{sec:res_123}Resonance across three minibands}

    \begin{figure}[b]                   
      \leavevmode
        \includegraphics[draft=false, height=4.8cm,angle=0,keepaspectratio=true]{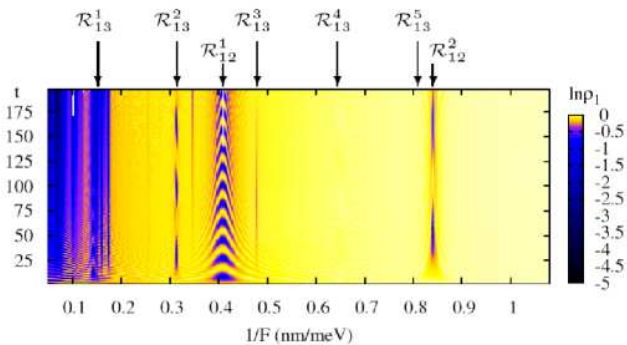}
      \caption[]{(Color online) Absolute occupancy function map for the first miniband in sample C; $\Psi(x,0)=w_1(x)$.}
      \label{fig:tanhHigh_W1_band1}
    \end{figure}

In a strong potential (sample C) having well-isolated minibands, we
were able to resolve $\R_{1\Btw3}^{2\ldots5}(C)$; the location of
several first resonances is indicated with longer arrows in
Fig.~\ref{fig:tanhHigh_W1_band1}. To our knowledge, the phenomenon
of Rabi oscillations across three minibands has been neither
observed nor simulated before.

\begin{figure}[t]               
        \includegraphics[draft=false,height=5.5cm,angle=0,keepaspectratio=true]{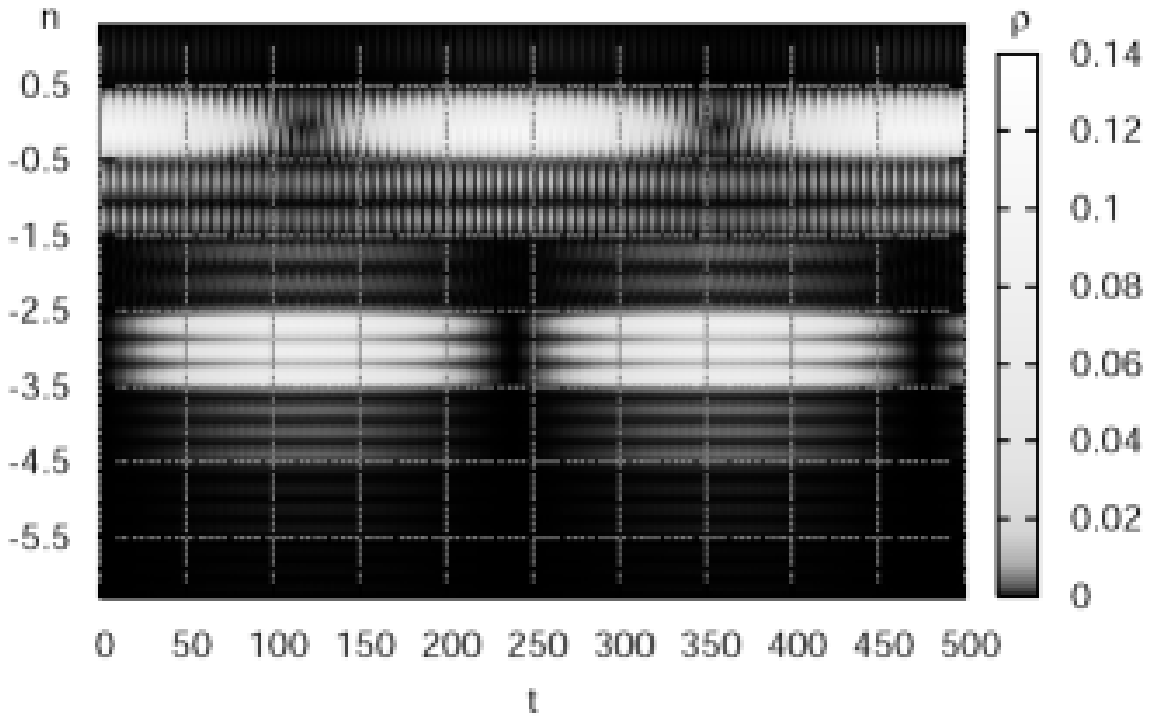}
        \includegraphics[draft=false,height=5.5cm,angle=0,keepaspectratio=true]{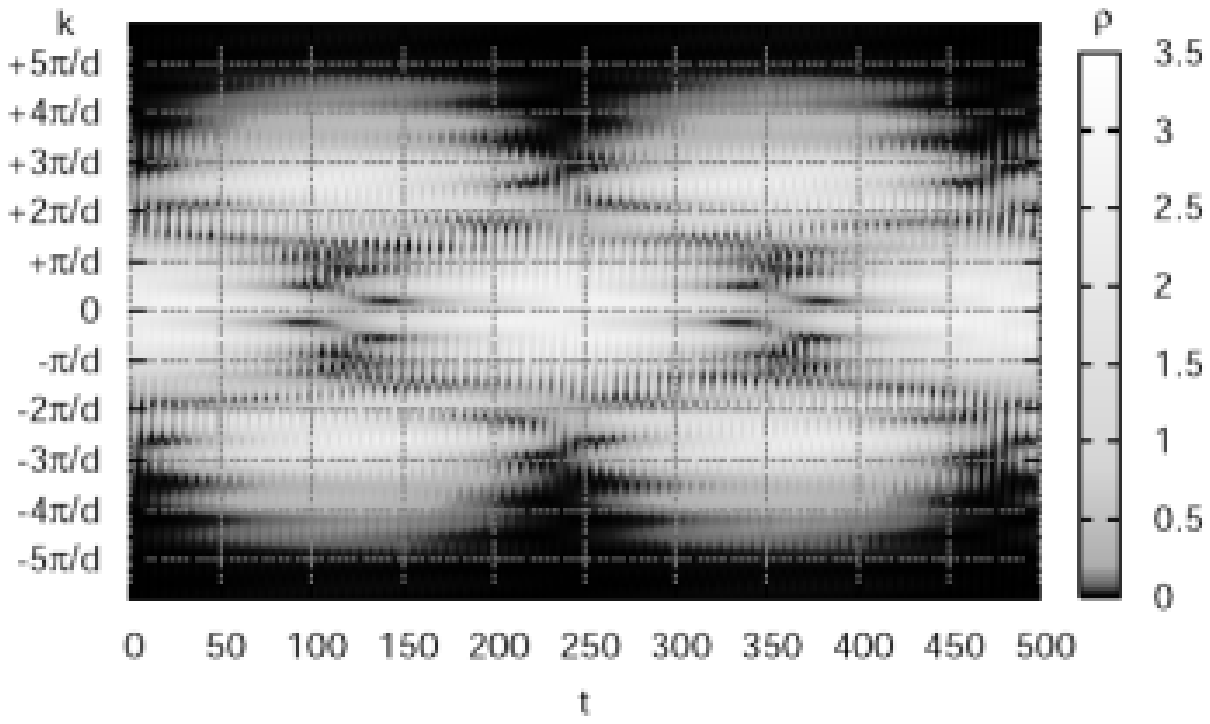}
      \caption[]{(Color online) Dynamics of the wavepacket $\Psi(x,0)=w_1(x)$ in real (top) and reciprocal (bottom) space at $\R_{1\Btw3}^3$ in sample C.}
      \label{fig:Rabi_res13_3_tanhHigh}
    \end{figure}

Dynamics of Rabi oscillations at $\R_{1\Btw3}^3$ shows that the
wavepacket resides mostly in the first and the third minibands. In
real space (upper panel of Fig.~\ref{fig:Rabi_res13_3_tanhHigh};
$\rho_{\nu}(t)$ is seen as a part of $\rho(x)$ whose shape has $\nu$
humps per cell), there is little probability residing in wells -1
and -2 at all times. In reciprocal space (lower panel), the second
Brillouin zone consistently remains underoccupied relative to the
first and third zones; some traces of $\rho(k,t)$ in the
fourth and the fifth Brillouin zones correspond to RZT, and the
finer background oscillations are caused by Bloch oscillations in
individual minibands. Thus, the carrier mostly bypasses the second
miniband since the resonance condition for $\R_{1\Btw2}$ is not well
satisfied, and tunnels directly into the third miniband.

It was found that $\R_{1\Btw2}$ and $\R_{1\Btw3}$ have many features
in common. Namely, the resonance shape was similar, the oscillations
of $\rho_1(t)$ were close to sinusoidal, Eq.~(\ref{eq:T_fit}) was
satisfied, and in k-space most part of a wavepacket
$\Psi(x,0)=w_1(x)$ traversed the first three Brillouin zones of the
TB model ($k\in[-\frac{3\pi}{d},\frac{3\pi}{d}]$) as a whole.
However, individual intrawell oscillations were not well pronounced
at $\R_{1\Btw3}$ because of the strong potential barriers in sample
C and the fact that intrawell oscillations between the pairs of
coupled minibands 1$\leftrightarrow$2 and 1$\leftrightarrow$3 are of
comparable magnitude and strongly interfere with each other, as was
noticed from the irregular shape of the resulting oscillations of a
wavepacket within the 0$^{th}$ cell. Also, within the 0$^{th}$ cell,
the center of the resultant oscillations of the probability density
was shifted down the potential ramp from the center of the cell,
under the influence of strong bias which creates asymmetry about the
cell center.


\subsection{\label{sec:res_123:sandwiched}Role of sandwiched miniband}

For a resonance across three minibands, an exponential fit based on
the two-miniband approximation is not a good fit to $T_n^{max}\,(n)$
any more (right panel of Fig.~\ref{fig:T_index_fit}). The reason lies
in the involvement of the second miniband which is ``sandwiched" between minibands 1 and
3, in the interminiband tunneling process. Despite its small average
population value, the second miniband must be taken into account at
any bias as will be shown below.

The role of the second miniband in carrier transfer between
minibands 1 and 3 must depend on resonance index, $n$. Otherwise,
$\langle\rho_2(t)\rangle$ should be proportional to the
interminiband tunneling rate of $\Psi(x,t)$, or inversely
proportional to the period of Rabi oscillations, and the latter
would be described by Eq.~(\ref{eq:fit_T_n}). Hence, this would
imply $\langle\rho_2\rangle\:\propto e^{-n}$.

    \begin{figure}[t]               
      \leavevmode
     \includegraphics[width=2.1cm,angle=270,keepaspectratio=true]{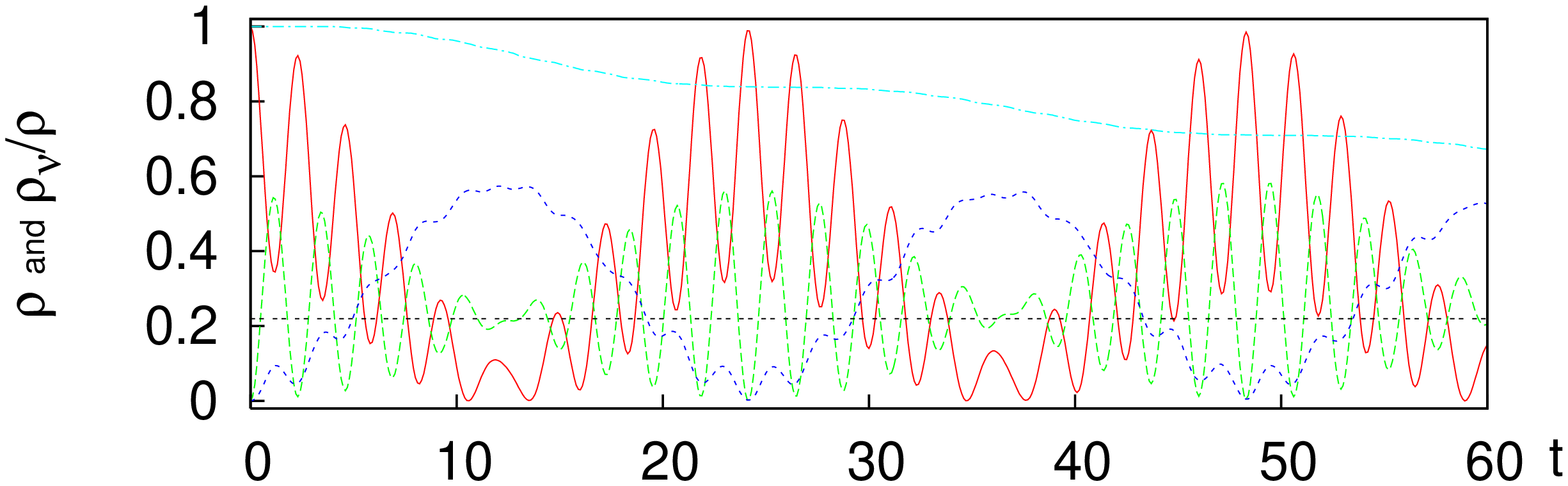}
     \includegraphics[width=2.1cm,angle=270,keepaspectratio=true]{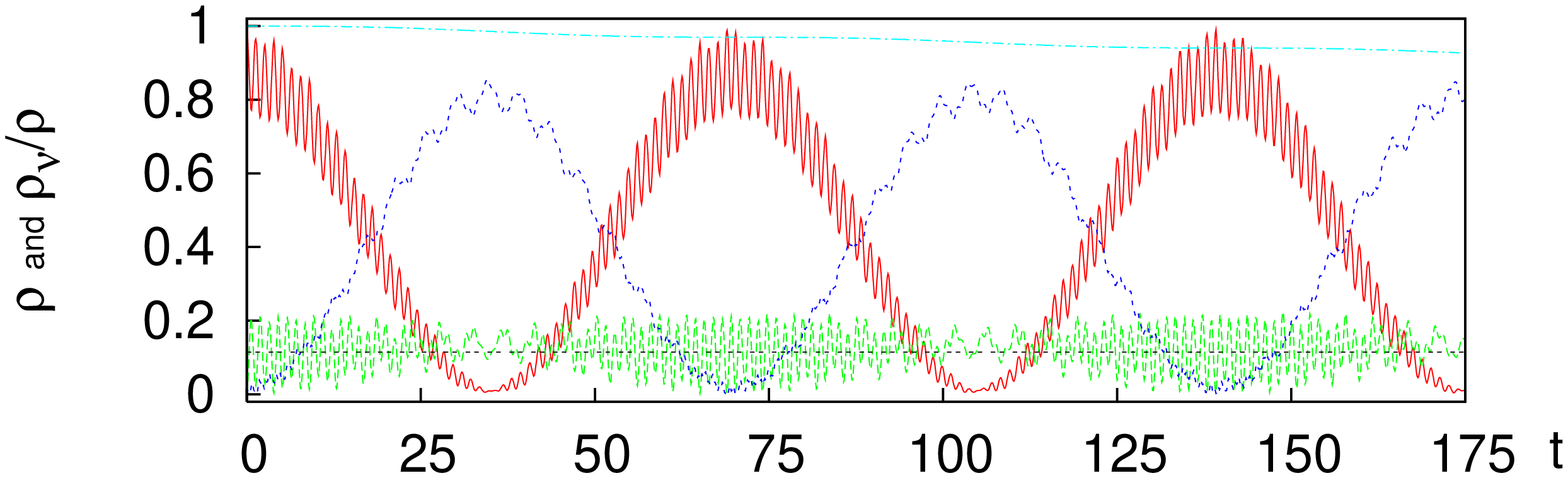}
     \includegraphics[width=2.1cm,angle=270,keepaspectratio=true]{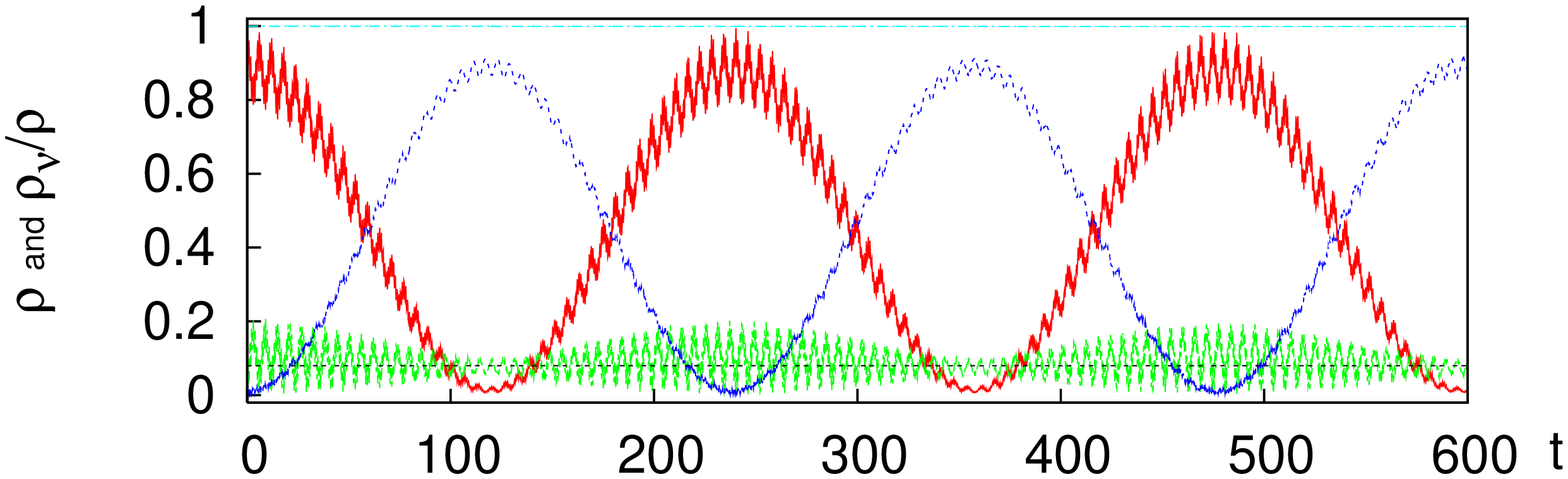}
     \includegraphics[width=2.1cm,angle=270,keepaspectratio=true]{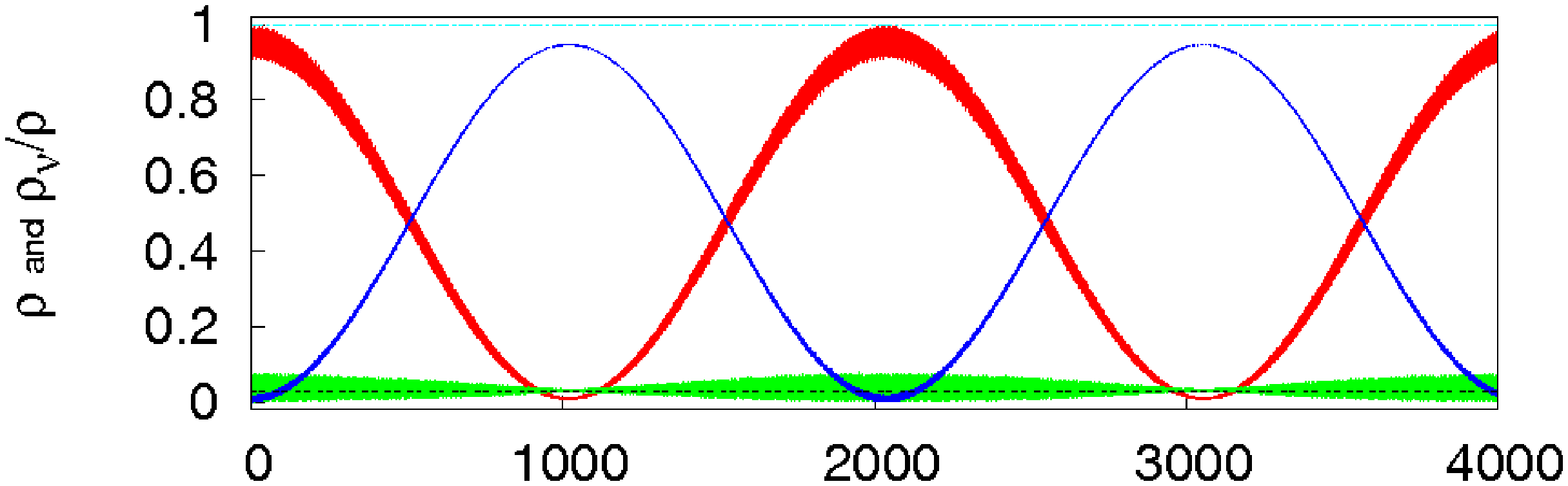}
     \includegraphics[width=2.1cm,angle=270,keepaspectratio=true]{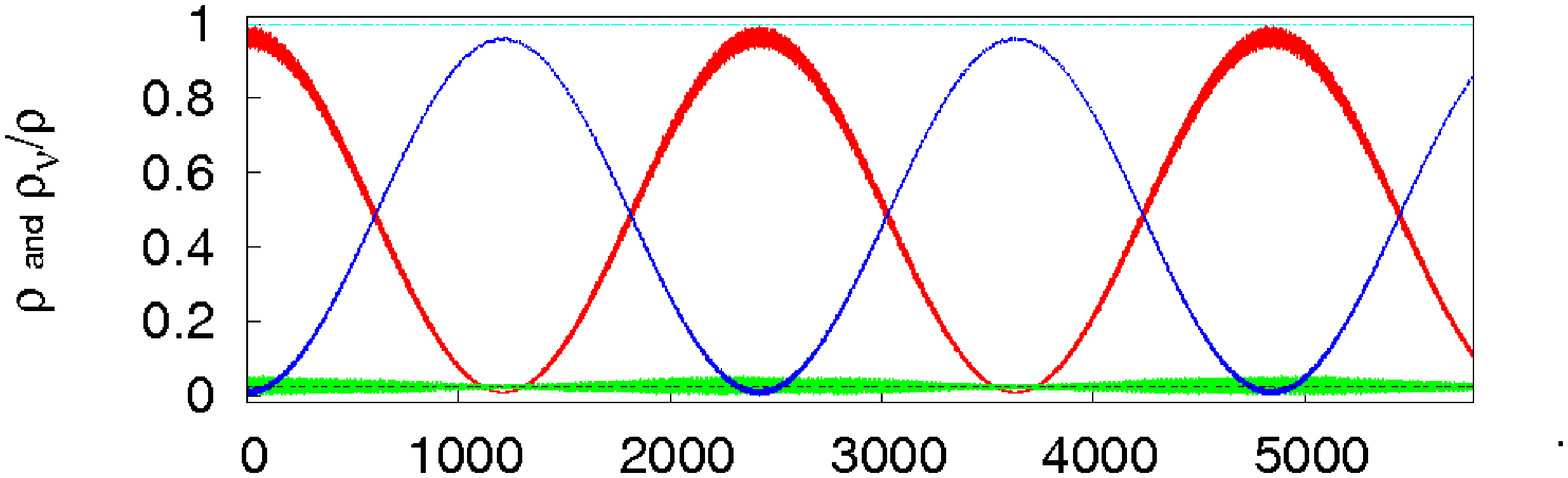}
      \caption[]{(Color online) Occupancy functions and wavepacket norm at $\R_{1\Btw3}^n(C)$
           for $n$ = 1 (top) through 5 (bottom). $\rho_1$ in solid (red), $\rho_2$ in dashed (green), $\rho_3$ in dotted
           (blue)and $\rho$ in chain-dotted (magenta) lines; the broken black line shows $\langle\rho_2\rangle$. }
      \label{fig:Rabi_at_res_13}
    \end{figure}

\begin{figure}[t]               
      \includegraphics[height=5cm,angle=0,keepaspectratio=true]{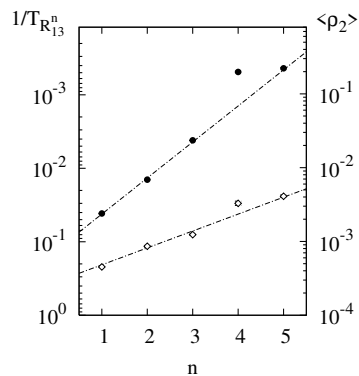}
      \caption[]{Rabi oscillation frequency (left scale, filled circles) and $\langle\rho_2\rangle$ (right scale, empty circles)
                 at $\R_{1\Btw3}(C)$ fitted with exponential curves (chain-dotted lines) on a logarithmic scale.}
      \label{fig:Mean_population_fit}
    \end{figure}

Values of $\langle\rho_2\rangle$ obtained are shown in
Fig.~\ref{fig:Rabi_at_res_13}, which displays dynamics of the
relative occupancy function at $\R_{1\Btw3}(C)$. It is remarkable
that, despite the strong coupling to the third miniband, the RZT
rate is small for the resonance indices $n>2$ thus allowing for
long-lasting Rabi oscillations. For lower resonance indices, the
population of the second miniband reaches significant level at
certain times. As the resonance index ascends, the average
population of the second miniband at $\R_{1\Btw3}^n$ decreases
(Fig.~\ref{fig:Rabi_at_res_13}); its population is largely caused by
Rabi oscillations between minibands 1 and 3. As
$\rho_2(t)\rightarrow0$, most of the wavepacket undergoes sinusoidal
oscillations between minibands 1 and 3, i.e. for lower biases the
interminiband dynamics resembles a two-miniband model (which would
assume $\rho_2(t)\equiv0$). The values of $\langle\rho_2\rangle$ are
shown separately in Fig.~\ref{fig:Mean_population_fit} and appear to
significantly deviate from exponential dependence at $\R_{1\Btw3}^4$.

In the process of Rabi oscillations at $\R_{1\Btw3}$, the population
level of the sandwiched second miniband is higher, the better is the
match in energy between the states from WSL2 and the initial carrier
energy $E_1^0$. The reason lies in the tunneling mechanism involved:
in the process of the multiwell tunneling between minibands 1 and 3,
an electron's energy after a certain number of interwell hops can be
close to $E_2$. This proximity greatly enhances population of the
sandwiched miniband by providing available density of states.
Conversely, a large population level of a sandwiched miniband favors
transition between the first and the third minibands. When the
transition element $x_{12}^{02}\,x_{23}^{25}$ becomes comparable to
$x_{13}^{05}$ (here $x_{\nu\mu}^{mn}\equiv\langle
w_{\nu}^n(x)|x|w_{\mu}^m(x)\rangle$), indirect tunneling through
the sandwiched miniband becomes significant. Therefore the degree of
participation of the second miniband and its effect on $\R_{1\Btw3}$
is determined by alignment of energy levels in WSL1, WSL2 and WSL3
and is dissimilar for resonances with different indices. The details
of the influence of this alignment on the electron dynamics will be
given in the next subsection \ref{sec:res_123:overlap}.

The correlation between coupling strength to the sandwiched miniband
and the rate of interminiband transition, or Rabi oscillation
frequency, is clearly demonstrated in
Fig.~\ref{fig:Mean_population_fit}: for $\R_{1\Btw3}^4$, both
$\frac{1}{T_{\R}}$ and $\langle\rho_2\rangle$ are smaller than
expected from the exponential fit. For index $n=4$, the tunneling
channel between the levels $E_1^0$ and $E_3^4$ consists of four
interwell hops, and the intermediate energy levels available to the
carrier are $E_2^{1,2}$. In this case, the alignment of $E_2^{1,2}$
turns out to be particularly unfavorable for them to act as a
strong transition channel. At $F=F_4=1.245~\units$, $E_2^1$ and
$E_2^2$ are equally remote from the carrier's initial energy:
$|E_1^0-E_2^1|\approx|E_1^0-E_2^2|$$\approx\frac{F_4d}{2}\approx20$~meV
(whereas in the other cases, when $n\neq4$ and $n>1$, one of the
levels $E_2^k$ is closer to $E_1^0$ than 7 meV). Such an alignment
can also be anticipated from Fig.~\ref{fig:tanhHigh_W1_band1}, where
$\R_{1\Btw3}^4$ lies in the middle between $\R_{1\Btw2}^1$ and
$\R_{1\Btw2}^2$ and hence is particularly isolated from any of the
$\R_{1\Btw2}^k$.

A sandwiched miniband affects not only Rabi oscillations, but also
RZT. In Fig.~\ref{fig:Toshima_W1_bands}, we can see that despite a
weaker bias, RZT at $\R_{1\Btw3}^5$ is stronger than for the
preceding $\R_{1\Btw3}^4$ for the same reason, namely proximity to
$\R_{1\Btw2}^2$. There, it also makes Rabi oscillations
corresponding to $\R_{1\Btw2}$ vanish at the given field since the
barrier through which the carrier tunnels to the third miniband and
then to the continuum is significantly reduced by coupling between
WSL2 and WSL3. In other words, closely situated resonances mutually
influence each other and can be referred to as coupled resonances.

    \begin{figure}[b]                   
      \leavevmode
        \includegraphics[draft=false, height=7.5cm,angle=0,keepaspectratio=true]{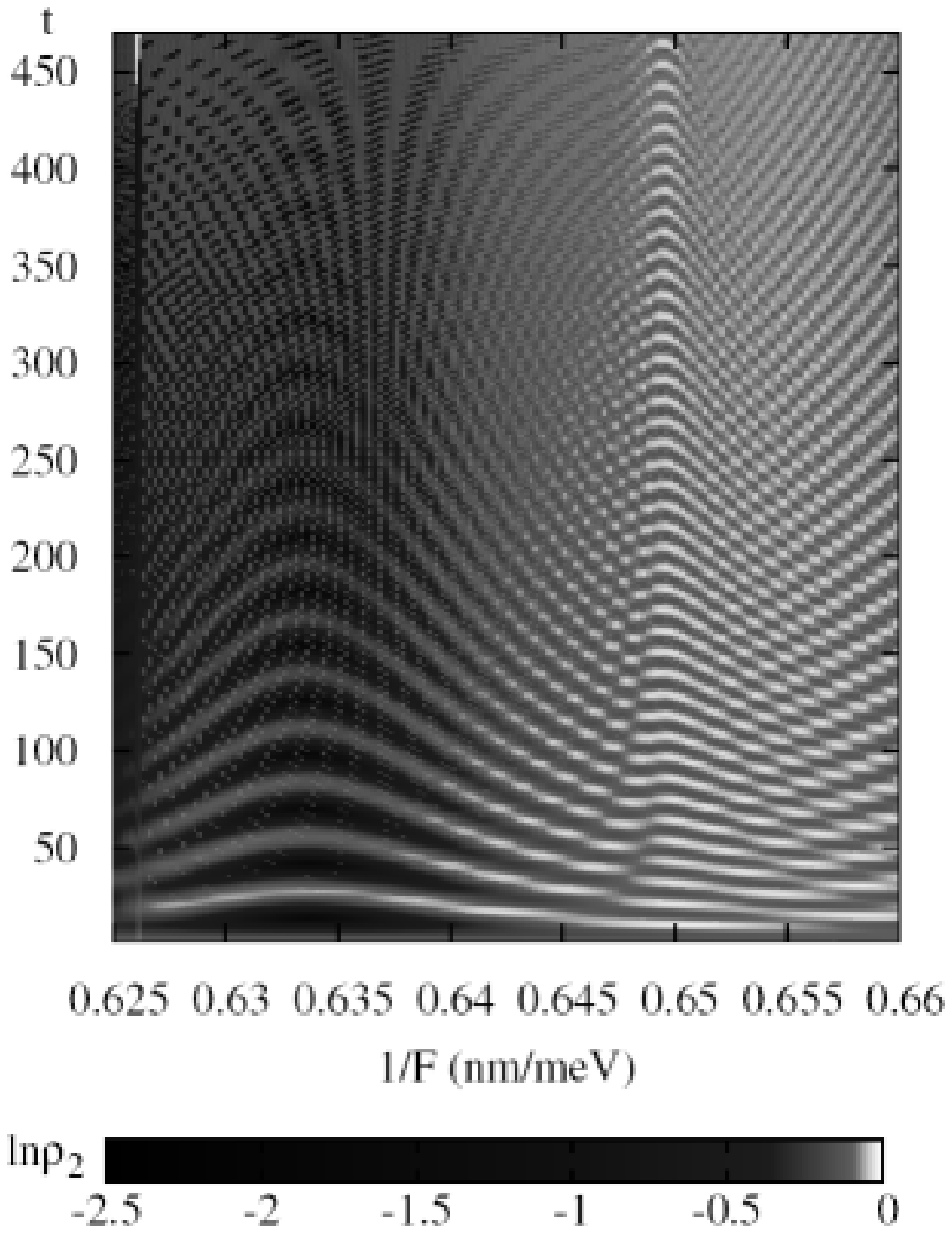}
        \includegraphics[height=8cm,angle=270,keepaspectratio=true]{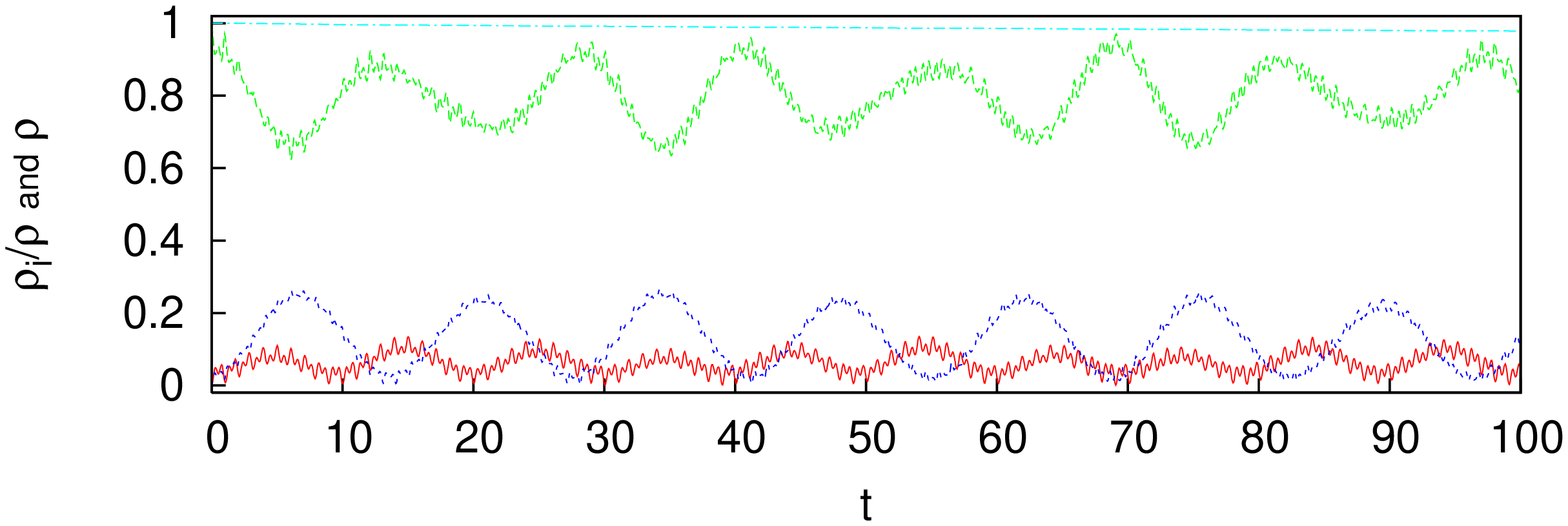}
      \caption[]{(Color online) Detailed view of coupled resonances ($\R_{1\Btw2}^2$,$\R_{2\Btw3}^3$ and $\R_{1\Btw3}^5$) in sample B for $\Psi(x,0)=w_2(x)$.
                 Relative occupancy functions dynamics at bias $\frac{1}{F}$=0.646~$\units$ is shown in the bottom panel; $\rho_1$ is shown in solid (red), $\rho_2$ in dashed (green), $\rho_3$ in dotted
                 (blue), and $\rho$ in chain-dotted (magenta) lines.}
      \label{fig:123_map}
    \end{figure}

Despite the sometimes small value of its average population, the
second miniband has a significant effect on resultant interminiband
dynamics. Thus to calculate parameters of a resonance across three
minibands, it is necessary to take at least these three minibands
into consideration.

\subsection{\label{sec:res_123:overlap}Coupled resonances}

In order to understand the interaction between different resonances
better, let us consider an example where strength of interminiband
coupling for different resonances is comparable and the resonances
strongly interfere with each other. The region corresponding to the
coupling of two resonances $\R_{2\Btw3}^3$ and $\R_{1\Btw2}^2$ in
sample B is shown in Fig.~\ref{fig:123_map}, a map plot of $\rho_2$
for $\Psi(x,0)=w_2(x)$. $\R_{1\Btw2}^2$ is situated on the left and
$\R_{2\Btw3}^3$ on the right, the former being wider due to its
lower resonance index.

    \begin{figure}[t]               
        \includegraphics[draft=false,height=5.5cm,angle=0,keepaspectratio=true]{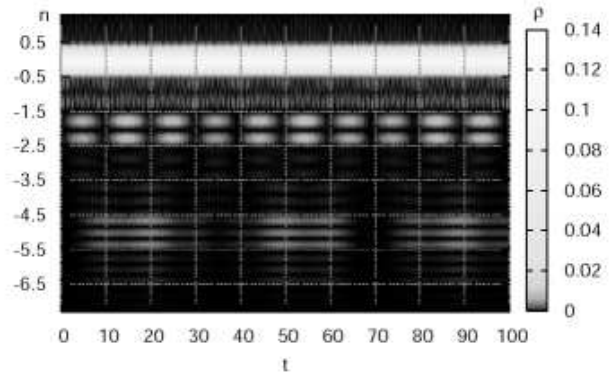}
      \caption[]{(Color online) Dynamics of the wavepacket $\Psi(x,0)=w_1(x)$ in direct space in the coupled resonances zone in sample B.}
      \label{fig:123_W1_tunneling}
    \end{figure}

Detail of the carrier dynamics for field
$\frac{1}{F}$=0.646~$\units$ where contributions from
$\R_{1\Btw2}^2$ and $\R_{2\Btw3}^3$ are nearly equal, are shown in
the bottom section of Fig.~\ref{fig:123_map}. Since all three
minibands are interacting at the same time, the resultant dipole
dynamics looks somewhat similar to a superposition of the three
corresponding Rabi oscillations for the individual resonances
($\R_{1\Btw2}^2$, $\R_{2\Btw3}^3$ and $\R_{1\Btw3}^5$), and the time
evolution of $\rho_{1,2,3}(t)$ exhibits beats. Therefore for coupled
resonances, the resultant Rabi oscillations are an interference
product of not only Bloch and intrawell oscillations, but also of
Rabi oscillations corresponding to the other pairs of strongly
coupled minibands in the sample. This interference allows the
carrier to use the ``extra" density of states to enhance tunneling.
A map plot of the wavefunction in the coupled resonances zone
(Fig.~\ref{fig:123_W1_tunneling}) demonstrates that large miniband
populations facilitate the build-up of each another. For example,
$\rho_2(t)$ is relatively low at $t=35~T_B$ and $t=70~T_B$, i.e. at
the moments when $\rho_3(t)$ is almost zero. Likewise when
$\rho_2(t)$ is relatively large, $\rho_3(t)$ becomes larger as well
(e.g. at $t=20~T_B$, 50~$T_B$ and 90~$T_B$).

Obviously the two-miniband model cannot reproduce such behavior. As
was recently found~\cite{Govorov05}, Rabi oscillations in a
three-level system are not merely a summation of Rabi oscillations
between the three separate pairs of minibands, they rather are a
coherent superposition. In fact, the temporary reduction of Rabi
oscillation amplitude observed in work~\cite{Zhao00} was due to Rabi
oscillation revival, i.e. the beats caused by superposition of Rabi
oscillations between several lowest minibands, rather than to Rabi
oscillation dephasing.


\section{Conclusion}

In conclusion, our time-dependent simulations of electron dynamics
in a biased superlattice provide an overview of carrier behavior
over a large range of bias. We were able to
identify energy level anticrossings as resonances displaying Rabi
oscillations and/or RZT and to study the structure of these
resonances in detail; our findings could be experimentally studied
by observing the system's pulsed output.

It has been shown that Rabi oscillations result from constructive
interference between Bloch and intrawell oscillations. A distinct
interminiband resonance occurs whenever the initial wavepacket is
capable of building up a Wannier-Stark state of significant
magnitude away from its initial location, through the coherent
process of self-interference.

We have also reported and analyzed Rabi oscillations across three
minibands. The important role of a sandwiched miniband in the
transitions across more than one miniband has been studied:
energy-wise it provides available density of states in the
intermediate tunneling region, and dynamically it invokes additional
Rabi oscillations interfering constructively with the principal
interminiband transition.

At a resonance across three minibands and in the coupled resonances
region, the resultant carrier dynamics has significant contribution
from all three minibands. Hence these cases cannot be described
within the commonly used two-miniband model, and require taking at
least three minibands into consideration.

Interminiband separations calculated in the tight-binding
approximation can be used to predict resonant values of bias in
the high-field regime with reasonable accuracy for a variety of
potentials, including resonances between minibands 2 and 3 and those
across three minibands (e.g. $\R_{1\Btw3}$, with the only
restriction $Fd<E_2-E_1$).

Although the present work focuses on an ideal zero-temperature
superlattice under uniform constant electric field, the numerical
methods utilized are easily capable of handling time-dependent
irregular potentials. With minor variations, the numerical
techniques can be applied to study carrier dynamics in other systems
(e.g. double quantum dots~\cite{Roskos92, Leo90} and irregularly
shaped potentials~\cite{Fibonacci96}) and even areas of physics,
such as photonics~\cite{Photonics1,Photonics2,Photonics3,Photonics4}
and cold atom optical traps~\cite{Optical1, Optical2, Optical3,
Optical4}; these are reserved for the future investigations.

\begin{acknowledgements}
This work was supported by NSERC-Canada Discovery Grant RGPIN-3198,
and was included in the M.Sc. thesis of the first author. We thank
Professor W. van Dijk for his great help with the algorithm
implementation and for valuable comments, and Christian Veenstra who
kindly contributed the initial code. The numerical simulations
presented were made on the facilities of the Shared Hierarchical
Academic Research Computing Network (SHARCNET: www.sharcnet.ca).
\end{acknowledgements}


\appendix
\section{Discrete transparent boundary conditions}
\label{appendix:TrBC}

Our numerical method was based on TrBC implemented with Numerov and
Crank-Nicholson methods for space and time respectively, with the
cumulative precision $\mathcal{O}((\delta x)^{5})$ in space and
$\mathcal{O}((\delta t)^{2})$ in time. We extended the discrete TrBC
as described in~\cite{Moyer} to the case of unequal saturation
potentials on opposite sides. For the most part, we keep the same
notation as in~\cite{Moyer}.

Let us express Eq.~(\ref{eq:Schrod}) in terms of finite differences.
We will start from the finite difference form of the system's
propagator which translates the time by $\delta t$
 \begin{eqnarray}
  \Psi(x,t+\Delta) &\approx& e^{-iH(t)\Delta}\,\Psi(x,t)   \nonumber
\end{eqnarray}
with $\Delta\,\equiv\,\frac{\delta t}{\hbar}$ and $H(t)$ being the
(possibly time-dependent) Hamiltonian of the system. Cayley's
approximation preserves unitarity and is exact to second order:
 \begin{eqnarray}
  e^{-iH\Delta} &=& \frac{1-\frac{1}{2}\,i\,H\Delta}{1+\frac{1}{2}\,i\,H\Delta}\ +\ \mathcal{O}\big(\Delta^3\big) \nonumber
\end{eqnarray}
A little algebra leads to the expression
 \begin{eqnarray}
  \bigg[\frac{\partial^2}{\partial x^2} \,-\, \frac{2\,mm^*}{\hbar^2} \Big( V(x,t)-\frac{2i}{\Delta} \Big)\bigg]\,
            \Big( \Psi(x,t+\Delta)+ \nonumber\\
            +\,\Psi(x,t) \Big)\ =\ \frac{8\,i\,mm^*}{\hbar^2\Delta}\,\Psi(x,t)\ +\ \mathcal{O}\big(\Delta^3\big)  \nonumber
\end{eqnarray}
with the notation $V(x,t)=V_{SL}(x)+xF$. It serves as a discretized
in time version of Eq.~(\ref{eq:Schrod}) and as a starting point for
discretization in space using the Numerov method.

In order to build the solution $\Psi(x,t)$ to Eq.~(\ref{eq:Schrod}),
we will use a uniform space grid consisting of $J$ points and
defined as $x_j=j\,\delta x,\ (j=0,1,\ldots,J-1)$ and uniform time
grid $t^{(k)}=k\,\delta t,\ (k=0,1,\ldots)$. The subscript $j$ will
refer to the inner points $x_j$, and $e$ to the end points $x_0$ or
$x_{J-1}$ as appropriate; the superscript $(k)$ will denote the time
instant $t^{(k)}$. Thus $\Psi(x_j,t^{(k)})\equiv\Psi_j^{(k)}$ and
$V(x_n)\equiv V_n$.

To apply TrBC in the Numerov approximation for the 1D case, the
necessary conditions are: $V(x<x_0)=V(x=x_0)\equiv V_0$, and
$V(x>x_{J-1})=V(x=x_{J-1})\equiv V_{J-1}$, as well as
$\Psi_e^{(0)}=0$ at both ends. Having started at $t=0$, we proceed
as follows to construct $\Psi(x,t^{(k+1)})$ from $\Psi(x,t^{(k)})$:

(i) Calculate the time-independent coefficients over the space grid:
 \begin{eqnarray}                           
  g_j      &=& \frac{2\,mm^*}{\hbar}\,(V_j - \frac{2\,i}{\Delta}) \ \ \ \ \ \ \  0 \leq j \leq J-1     \nonumber\\
  d_j      &=& 1 - \frac{(\delta x)^2}{12}\,g_j  \ \ \ \ \ \ \ \ \ \ \ \ \ \ 0 \leq j \leq J-1                     \nonumber\\
  e_j      &=&  \left\{\begin{array}{ll}
                \alpha_0                                               & \textrm{$j=0$} \\
                2 + (\delta x)^2\, \frac{g_j}{d_j} - \frac{1}{e_{j-1}}     & \textrm{$0<j<J-1$}
                      \end{array}
                \right.                                  \nonumber
\end{eqnarray}

(ii) Calculate the time-independent border coefficients ($e = 0$ or
$J-1$ separately):
 \begin{eqnarray}                           
  a_e      &=& 1 + \frac{(\delta x)^2}{2}\,\frac{g_e}{d_e}               \nonumber\\
  \alpha_e &=& a_e + \sqrt{a_e^2-1}                                      \nonumber\\
  c_e      &=& 1 - \frac{2\,i\,(\delta x)^2 mm^*}{3\delta t}\,\frac{1}{d_e}  \nonumber\\
  \phi_e   &=& \arg\Big( \frac{a_e^2 - 1}{c_e} \Big)                     \nonumber\\
  A_e      &=& \frac{1-|a_e|^2}{|1-a_e^2|}                               \nonumber\\
  \sigma_e &=& d_e(a_e-\alpha_e)                                     \nonumber\\
  \rho_e   &=& d_e^*(a_e^*-\alpha_e)                                 \nonumber
\end{eqnarray}

(iii) Construct the polynomials for the next time step:
 \begin{eqnarray}                           
  P_e^{(k)}    &=&  \left\{\begin{array}{ll}
                 1                                           & \textrm{$k=-1$} \\
                 A_e                                         & \textrm{$k=0$}  \\
                 \frac{2k+1}{k+1}\,A_e P_e^{(k-1)} - \frac{k}{k+1}\,P_e^{(k-2)}  & \textrm{$k>0$}
                       \end{array}
                \right.                                                          \nonumber\\
                                               \nonumber\\
  L_e^{(k)}    &=&  \left\{\begin{array}{ll}
                e^{-i\phi_e}\,P_e^{(k)}                                                       & \textrm{$k=0$} \\
                \frac{1}{2k+1}\,e^{-i(k+1)\phi_e}\Big(P_e^{(k)}-P_e^{(k-2)}\Big)   & \textrm{$k>0$}
                      \end{array}
                \right.                                  \nonumber
\end{eqnarray}

(iv) Calculate the time-dependent coefficients over the space grid:
 \begin{widetext}
 \begin{equation}                   
  q_j^{(k+1)}  =  \left\{\begin{array}{ll}
                \rho_0\Psi_0^{(k)} + \sigma_0 \sum^{k}_{m=0}\, L_0^{(k-m)}\Psi_0^{(m)}    & \textrm{$j=0$} \\
                \frac{q_{j-1}}{e_{j-1}} + \frac{\zeta}{d_j}\Psi_j^{(k)}             & \textrm{$0\leq j<J-1$}
                      \end{array}
                \right.                                  \nonumber
\end{equation}
 \begin{equation}                   
  w_j^{(k+1)}  =  \left\{\begin{array}{ll}
        \Big[ q_{J-2}^{(k+1)}+e_{J-2}\,\Big(\rho_{J-1}\Psi_{J-1}^{(k)} + \sigma_{J-1} \sum^{k}_{m=0}\, L_{J-1}^{(k-m)}\Psi_{J-1}^{(m)} \Big) \Big]
                  \Big(1-\alpha_{J-1} e_{J-2}\Big)^{-1}                      & \textrm{$j=J-1$} \\
        \frac{1}{e_j}({w_{j+1}^{(k+1)} - q_j^{(k+1)}})             & \textrm{$0<j<J-1$}
                      \end{array}
                \right.                                  \nonumber
\end{equation}
\end{widetext}

(v) Based on the above, construct wavefunction for the next instant
of time:
 \begin{eqnarray}                   
  \Psi_j^{(k+1)} &=& \frac{w_j^{(k+1)}}{d_j} + \big(\frac{\xi}{d_j}-1 \big)\Psi_j^{(k)}           \nonumber
\end{eqnarray}
with the notation
 \begin{eqnarray}                   
  \zeta         &\equiv& \frac{8\,i\,(\delta x)^2\, mm^*}{\delta t}                 \nonumber\\
  \xi           &\equiv& \frac{2\,i\,(\delta x)^2\, mm^*}{3\,\delta t}              \nonumber\\
  \Delta        &\equiv& \frac{\delta t}{\hbar}                                     \nonumber
\end{eqnarray}
and $P_e^{(k)}$ being Legendre polynomials of the $k^{th}$ order
having $A_e$ at the time instant $t=t^{(k)}$ as an argument.

If the potential considered is time-dependent in the inner region,
the coefficients from step (ii) will change with time, and we will
simply have to recalculate them for every new instant of time. In
the formula language, this means substitution everywhere above
$V(x_j)\rightarrow V(x_j, t^{(k)})\equiv V_j^{(k)}$, which results
in time-dependent coefficients $g_j\rightarrow g_j^{(k)}$,
$d_j\rightarrow d_j^{(k)}$, $e_j\rightarrow e_j^{(k)}$, etc.



\end{document}